\documentclass[10pt,journal,compsoc,twoside]{IEEEtran}
\pdfoutput=1 

\usepackage[utf8]{inputenc}
\usepackage[english]{babel}
\usepackage[dvipsnames]{xcolor}

\usepackage[nocompress]{cite}
\usepackage{url}
\usepackage{hyperref}
\hypersetup{
    linkcolor  = violet!85!black,
    citecolor  = magenta!85!black,
    urlcolor   = blue!85!black,
    colorlinks = true,
    breaklinks = true
}

\usepackage{multirow, multicol, booktabs, tabu, array}
\usepackage{amsmath, amsfonts, amssymb, amsthm, nicefrac}
\DeclareMathOperator*{\argmin}{arg\,min}
\renewcommand{\footnotesize}{\scriptsize}
\theoremstyle{definition}
\newtheorem{definition}{Definition}
\newtheorem{lemma}{Lemma}
\usepackage[ruled]{algorithm2e}
\usepackage{algorithmic}

\usepackage[inline]{enumitem}

\usepackage[pdftex]{graphicx}
\usepackage{url}
\usepackage{epstopdf, subfigure, stfloats, bbding, capt-of}
\usepackage{balance}
\graphicspath{{figs/}}
\usepackage{cleveref} 
\crefformat{section}{Section~#2#1#3}
\crefmultiformat{section}{Sections~#2#1#3}{ and~#2#1#3}{, #2#1#3}{ and~#2#1#3}
\crefformat{figure}{Fig.~#2#1#3}
\crefformat{table}{Table~#2#1#3}
\crefformat{equation}{(#2#1#3)}
\crefformat{algorithm}{Algorithm~#2#1#3}
\crefformat{definition}{Definition~#2#1#3}
\crefformat{lemma}{Lemma~#2#1#3}
\usepackage{soul}

\usepackage[final]{microtype}
\makeatletter
\newcommand\subparagraph{%
	\@startsection{subparagraph}{5}
	{\parindent}
	{3.25ex \@plus 1ex \@minus .2ex}
	{-1em}
	{\normalfont\normalsize\bfseries}}
\makeatother
\usepackage[compact]{titlesec}
\let\subparagraph\relax 
\titlespacing\section{0pt}{6pt plus 4pt minus 2pt}{2pt plus 2pt minus 2pt}
\titlespacing{\subsection}{0pt}{4pt plus 2pt minus 1pt}{2pt plus 1pt minus 1pt}
\titlespacing{\subsubsection}{0pt}{4pt plus 2pt minus 1pt}{2pt plus 1pt minus 1pt}
\usepackage{etoolbox}
\makeatletter
\patchcmd{\ttlh@hang}{\parindent\z@}{\parindent\z@\leavevmode}{}{}
\patchcmd{\ttlh@hang}{\noindent}{}{}{}
\makeatother

\begin{document}
    \title{ASAP: \underline{A}ccelerated \underline{S}hort-Read \underline{A}lignment on 
    \underline{P}rogrammable Hardware}
    \author{Subho~S.~Banerjee, 
        Mohamed~El-Hadedy, 
        Jong~Bin~Lim, \\
        Zbigniew T. Kalbarczyk, 
        Deming Chen, 
        Steven S. Lumetta, 
        and~Ravishankar~K.~Iyer
        \IEEEcompsocitemizethanks{\IEEEcompsocthanksitem S. S. Banerjee, M. el-Hadedy, J. B. Lim,
        	Z. T. Kalbarczyk, D. Chen, S. Lumetta and R. K. Iyer are with the Coordinated 
        	Science Laboratory, and the Departments of Computer Science and Electrical and Computer 
        	Engineering all at the University of Illinois at Urbana-Champaign, Urbana,	IL, 61801.}%
    }


    \makeatletter
    \def\ps@IEEEtitlepagestyle{%
        \def\@oddfoot{\mycopyrightnotice}%
        \def\@evenfoot{}%
    }
    \def\mycopyrightnotice{%
        {\footnotesize This work has been submitted to the IEEE for possible publication. Copyright
    may be transferred without notice, after which this version may no longer be
    accessible.\hfill}%
        \gdef\mycopyrightnotice{}
    }

    \IEEEtitleabstractindextext{

\begin{abstract}
The proliferation of high-throughput sequencing machines ensures rapid generation of up to billions
of short nucleotide fragments in a short period of time. This massive amount of sequence data can
quickly overwhelm today’s storage and compute infrastructure. This paper explores the use of
hardware acceleration to significantly improve the runtime of short-read alignment, a crucial step
in preprocessing sequenced genomes. We focus on the Levenshtein distance (edit-distance) computation
kernel and propose the ASAP accelerator, which utilizes the intrinsic delay of circuits for
edit-distance computation elements as a proxy for computation. Our design is implemented on an
Xilinx Virtex 7 FPGA in an IBM POWER8 system that uses the CAPI interface for cache coherence across
the CPU and FPGA. Our design is $200\times$ faster than the equivalent C implementation of the
kernel running on the host processor and $2\times$ faster for an end-to-end alignment tool for
120--150 base-pair short-read sequences. Further the design represents a $3760\times$ improvement
over the CPU in performance/Watt terms.
\end{abstract}

\begin{IEEEkeywords}
    Bioinformatics, Genomics, Levenshtein Distance, Application-Specific Processor, Hardware 
Accelerator.
\end{IEEEkeywords}}

    \maketitle
    
    \section{Introduction} \label{sec:introduction}
\IEEEPARstart{T}{he} advent of high-throughput next-generation sequencing technology (NGS) has
created a deluge of genomic data for computational analysis~\cite{Stephens15}. Efficiently
processing this data requires the development of a new generation of high-performance computing
systems that can efficiently handle such data. This new generation of application-specific and
accelerator-rich computing systems are expected to gain performance, power, and energy improvements
over traditional systems~\cite{Shao2015}.

A crucial step in a significant number of NGS data analytics applications (e.g., variant discovery,
genome-wide association studies, and phylogeny creation) is the mapping of short fragments of
sequenced genetic material (called \emph{reads}) to their most likely points of origin in the
genome, popularly called the \emph{short-read alignment} problem. This paper presents the design and
implementation of ASAP, an accelerator for computing Levenshtein distance~\cite{Levenshtein66,
Navarro2001} (LD; used interchangeably with edit-distance) in the context of the short-read
alignment problem. LD is a measure of the similarity between strings, which is computed by counting
the number of single-character edits required to change one string into the other.  LD computation
is a prominent underlying mathematical kernel that is common to a large number of short-read
alignment algorithms and tools (e.g., BLAST~\cite{Altschul1990}, Bowtie~\cite{Langmead09,
Langmead12}, BWA~\cite{Li10}, and SNAP~\cite{Zaharia11}), and is responsible for 50\% -- 70\% of
their runtime~\cite{Banerjee2016}.

ASAP represents a novel approach to accelerate the LD computation, in that it uses algorithmic
approximations, and maps these approximations into hardware to significantly improve overall
performance ($\sim 200 \times$ compared to the CPU baseline).  The core algorithm in ASAP leverages
two key observations about the computation and datasets involved in the short-read alignment
problem:
\begin{enumerate}[noitemsep,nolistsep,leftmargin=*]
    \item \label{item:approximation_idea} Although all the tools mentioned above calculate the exact
    value of LD between pairs of nucleotide strings, they use them only \emph{to build a total
    ordering} (i.e., an ordered list) of the most likely points of origin in the genome. The best 
    alignment is the pair of strings corresponding to the minimum LD in the
    ordered list. Hence, it is sufficient to only calculate the total ordering (in this instance,
    returning the pair that corresponds to the minimum LD), and not essential to compute the exact
    value of the LD. This distinction enables approximation in the computation of LD to gain
    performance, while preserving the overall accuracy of the alignment algorithm (which comes from
    the total ordering).
    \item \label{item:data_matches} Modern sequencing platforms (like the Illumina HiSeq 2500)
    represent a very low sequencing error regime ($\leq 1 \%$)~\cite{Glenn2011, Ross2013}, and
    modern alignment tools (mentioned above) have accurate candidate region-matching algorithms
    (described in \cref{sec:ld_and_alignment}).  Hence, LD computations process 
    significantly more ``matches'' than ``mismatches,'' in the majority of sequencing
    experiments.\footnote{This is a facet of the accurate sequencing process and the thoroughly
    validated reference genome for human subjects. This observation will also apply to most model
    organisms whose genome has been extensively studied.} The ASAP architecture uses this 
    heuristic to accelerate LD computation (described in 
    \cref{sec:asap_delay_idea,sec:asap_approximation}).
\end{enumerate}
To take advantage of these observations, ASAP augments
RaceLogic~\cite{Madhavan2014}\footnote{RaceLogic uses propagation delay of circuit elements to
perform computations.} using application heuristics, as well as hardware architectural optimizations
to realize the design on FPGAs. In particular, this paper proposes
\begin{enumerate*}[label=(\alph*)]
    \item a mechanism to encode LD computation parameters (e.g., \emph{gap-penalties}; described 
    further in \cref{sec:ld_and_alignment}) into the ASAP architecture, making it possible to
    map the time taken to process a ``match'' exactly as a circuit delay. This mapping gives us the 
    ability to tune the performance of ASAP to match data characteristics; and
    \item the use of ``zero delay'' circuit elements to explore large portions of the search space 
    (LDs of substrings of the strings being compared) in parallel within one clock cycle, and to 
    ignore parts of the search space that do not contribute to an answer, thereby saving energy.
\end{enumerate*}
Overall, ASAP can compute alignments quickly ($\sim 200 \times$ faster than the CPU baseline
and $\sim 50 \times$ faster than an equivalent RaceLogic design), and with the same accuracy as
traditional software- or hardware-based alignment tools. We leverage reconfigurable
FPGA devices to prototype ASAP, thereby allowing us to reconfigure the accelerator based on user
decisions on input parameters (described in \cref{sec:ld_and_alignment}), as well as to adapt
the accelerator to input NGS datasets of varying read lengths.

\textbf{Contributions.} To summarize, the primary contributions of this paper are as follows:
\begin{enumerate}[noitemsep,nolistsep,leftmargin=*]
    \item Presents a measurement-driven study that demonstrates that computation of LD represents a
    significant portion of the runtime of several short-read alignment programs.
    \item Builds on top of the delay-based computation paradigm presented in
    \cite{Madhavan2014} to encode gap-penalties as ``zero delay'' circuit elements. This allows us
    to calculate approximate the LD between strings by using combinational circuit elements. We
    prove the correctness of this encoding and demonstrate that the result of the approximation can
    be used as a proxy for computing LD in short-read aligners. That is, a tool using the
    approximation and the accelerator produces alignments identical to those of tools based on
    traditional methods (e.g., BWA-MEM~\cite{Li10}).
    \item Presents an FPGA-based implementation of the accelerated LD computation in the ASAP
    accelerator that leverages the coherent accelerator-processor interface
    (CAPI)~\cite{CAPI_UserManual2014, CAPI2015} for communication between the host and accelerator.
    \item Demonstrates that ASAP on an FPGA is able to accelerate the runtime of the LD computation
    by $200\times$ compared to a CPU-based execution, and by $\sim5\times$ over the best xFPGA
    result, while consuming less energy.
    \item Demonstrates that integration of the ASAP accelerator into a short-read alignment
    framework like SNAP can accelerate it by nearly $2 \times$ (which is close to the Amdahl's
    law limit for the accelerator).
\end{enumerate}

\textbf{Other Applications.} Our approach can be adapted to a variety of other problems in which a
total ordering of LDs is computed. For example, in signal processing, where similarity between
signals is computed~\cite{Levenshtein66}; in text retrieval, where misspelled words have to be
accounted for in a dictionary~\cite{Baeza-Yates1999}; and in computer-security where virus- and
intrusion-detection requires comparison of signatures~\cite{Kumar94}.

\textbf{Organization.} The remainder of this paper is organized as follows.
\cref{sec:ld_and_alignment} describes the recursion-based formulation of the LD computation and its
use in popular short-read alignment tools. \cref{sec:asap} briefly describes
\begin{enumerate*}
    \item a mathematical formalism for encoding computation in circuit delays; 
    \item uses that formalism to define the approximation algorithm at the core of ASAP and prove 
    its correctness; and 
    \item presents the hardware architecture of ASAP leverages this approximation algorithm.
\end{enumerate*}
\cref{sec:evaluation} presents the evaluation of the accelerator. \cref{sec:related_work} compares
the ASAP approach to other CPU-, GPU-, and FPGA-based approaches for computing LD, and, finally, we
conclude in \cref{sec:conclusion}.
    
    \section{Levenshtein Distance Computation and Short-Read Alignment} \label{sec:ld_and_alignment}
Traditional methods for aligning reads to a reference genome find the position (\emph{locus}) of a
single read in the reference by minimizing the maximum edit distance between the short read being
aligned (called the \emph{query}, and denoted by $Q$) and the reference genome sequence. The
Smith-Waterman algorithm (SW)~\cite{Smith81} and Needleman-Wunsch algorithm
(NW)~\cite{Needleman1970} utilize a dynamic programming-based algorithm to calculate the alignment
score (Levenshtein distance) between the read and a particular section~$R$ of the reference genome,
accounting for base pair substitutions, insertions, and deletions. Both of these algorithms work by
constructing a matrix $S$ (which is used interchangeably with lattice) of size $l_Q \times l_R$,
where $l_Q$ and $l_R$ are the lengths of the two strings, between which the edit distance must be
calculated. Consider a matrix $S$ in which the $(i,j)^{th}$ entry, $S(i, j)$, is the minimum edit
distance between the sub strings $Q[1:j]$ and $R[1:i]$. $S(i,j)$ is recursively defined as
\begin{equation}
  \label{eqn:basic_levenshtein_distance}
  S(i,j) = min\left\{
          \begin{array}{ll}
            S(i-1, j) + \Delta(-, R_j),\\
            S(i, j-1) + \Delta(Q_i, -),\\
            S(i-1, j-1) + \Delta(Q_i, R_j)
          \end{array}
        \right\}
\end{equation}
where $\Delta$ corresponds to input parameters called \emph{gap penalties}. These
$\Delta$-parameters assign scores for insertion, deletion, match,\footnote{Gap penalties
traditionally do not have match scores. We group them together for simplicity in our notation.} or
mismatch between the sequences such that a more desirable outcome has a smaller score associated
with it. The parameters $\Delta(Q_i, R_j)$, $\Delta(-, R_j)$ and $\Delta(Q_i, -)$ correspond to the
match/mismatch, deletion, and insertion penalties respectively. These parameters are chosen to
optimize the accuracy of alignments based on prior information about the sequences being compared
(e.g., evolutionary information about mutations in a population~\cite{Sung2009, Henikoff1992,
Wang2011}). This paper describes the use of constant gap penalties (i.e., a fixed score is assigned
to every gap between nucleotides). That is,
\begin{equation}
    \left.
    \begin{array}{rcl}
        \Delta(Q_i, R_j) &=& \Delta(\text{Match}) \text{ if } Q_i=R_j\\
        \Delta(Q_i, R_j) &=& \Delta(\text{Mismatch}) \text{ if } Q_i \neq R_j\\
        \Delta(-, R_j) &=& \Delta(\text{Delete}) \\
        \Delta(Q_i, -) &=& \Delta(\text{Insert})
    \end{array}
    \right\}~\forall R_i, Q_j.
\end{equation}
Such gap penalties are are commonly used in DNA alignment (e.g., in NCBI-BLASTN, or
WU-BLASTN~\cite{Sung2009}).

\begin{figure}[!t]
    \centering
    \includegraphics[width=\columnwidth]{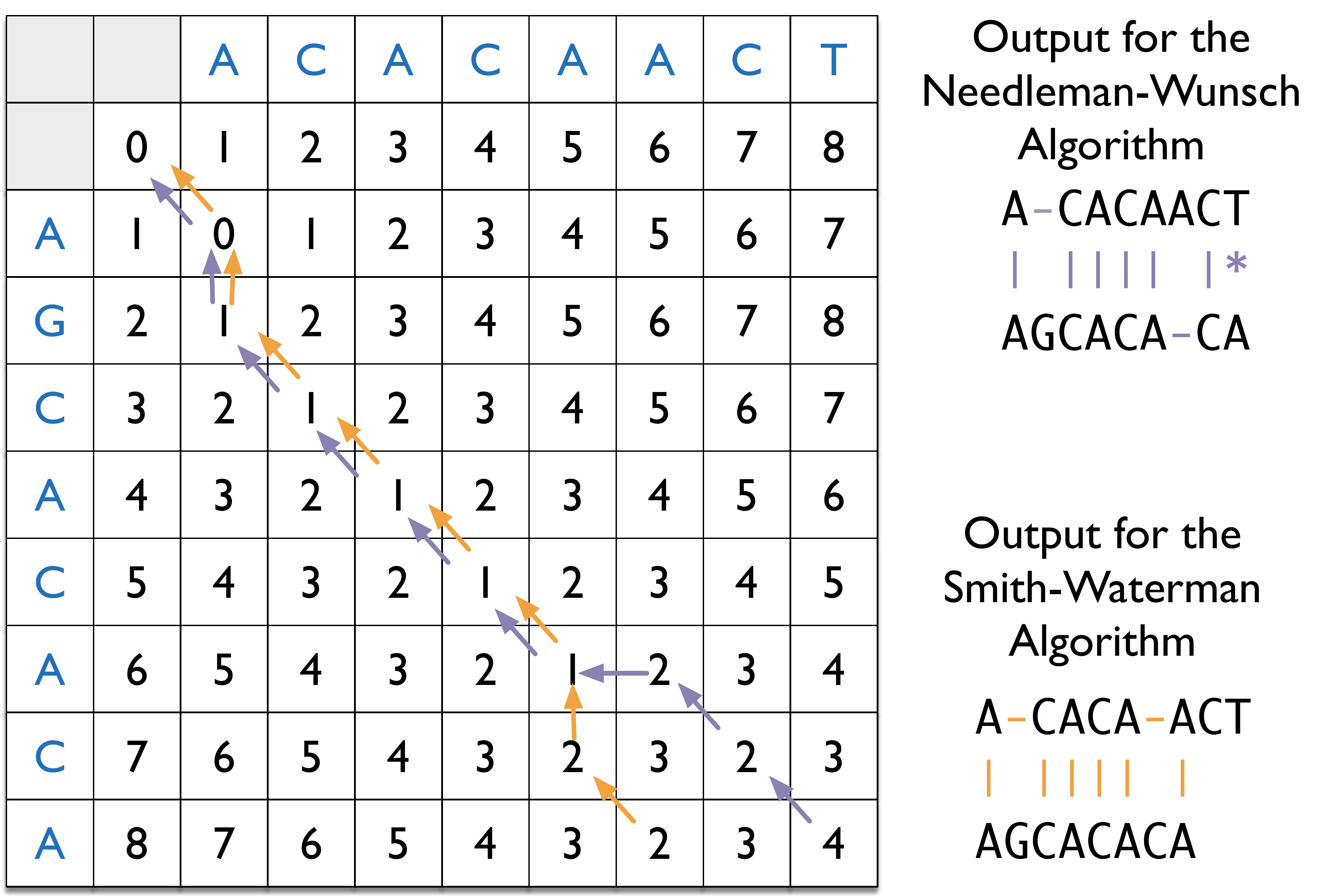}
    \caption{The matrix $S$ for the strings \texttt{AGCACACA} and \texttt{ACACAACT}, assuming 
        $\Delta(\text{Match}) = 0$, $\Delta(\text{Mismatch}) = 2$, and $\Delta(\text{Insert}) = 
        \Delta(\text{Delete}) = 1$. The colored paths from $S(8,8)$ and
        $S(8, 6)$ to $S(0, 0)$ show the optimal alignments produced by the NW and SW algorithms, 
        respectively.}
    \label{fig:dynamic_prog_matrix}
\end{figure}

The NW algorithm computes a global alignment in which the entirety of the query is matched to the
reference, as shown in \cref{fig:dynamic_prog_matrix}. It does so by computing the value of
$S(m, n)$. The SW algorithm computes a local alignment and matches the largest (substring) of the
query to the reference, and, hence, needs to calculate the minimum value in the row $S(m, -)$. For
example, when the strings \texttt{AGCACACA} and \texttt{ACACAACT} are compared with constant
penalties $\Delta(\text{Match}) = 0$, $\Delta(\text{Mismatch}) = 2$, and $\Delta(\text{Insert}) =
\Delta(\text{Delete}) = 1$, we get the matrix described in \cref{fig:dynamic_prog_matrix}.
The optimal alignment is then calculated from this matrix by finding the minimum weighted path (in
$S$) from $(m,n)$ to $(0,0)$ in the NW algorithm and $(m, \mathcal{N})$ to $(0,0)$ in the SW
algorithm. $\mathcal{N}$ corresponds to the largest substring of the reference to which the query
string maps with the lowest LD.

\begin{table*}[!t]
    \centering
    \caption{Mathematical formulation of different aligners to fit them into the structure of 
        \cref{algo:basic_alignment_skeleton}.}
    \label{tab:alignment_pieces}
    \begin{tabular}{lll}
        \toprule
        \textbf{Function} & \textbf{BWA-MEM}~\cite{Li10} & \textbf{SNAP}~\cite{Zaharia11}\\
        \midrule
        \texttt{Build\_Index}& Burrows-Wheeler transform~\cite{Burrows1994} of prefix trie& 
        Ukkonen's
        algorithm~\cite{Ukkonen1985} \\ 
        \texttt{Candidate\_Locations} & Prefix trie traversal & Hash table lookup\\
        \texttt{Edit\_Distance}& {Smith-Waterman algorithm~\cite{Smith81}} & 
        {Landau-Vishkin algorithm~\cite{Landau1986}}\\
        \texttt{Find\_Config} & {Smith-Waterman algorithm~\cite{Smith81}} & {Landau-Vishkin 
            algorithm~\cite{Landau1986}}\\
        \bottomrule
    \end{tabular}
    \vspace{-10pt}
\end{table*}

Although these methods are guaranteed to produce the optimal alignment, they are prohibitively
expensive for whole-genome alignments because of $O(l_Q \times l_R)$ space and time complexity.
Therefore, a large number of alignment tools are designed to heuristically reduce the search space
required to find the optimal match of a query in the reference. An extensive amount of research,
e.g.,~\cite{Altschul1990, Langmead09, Langmead12, Li10, Zaharia11}, has been conducted, focusing on
indexing strategies for the reference genome to rapidly reduce the number of candidate locations
that have to be searched.  Most of these tools use some variant of a backwards search algorithm
utilizing an FM-index~\cite{Ferragina2000} or a hash-table-like data structure. As a result of this
reduction in the search space, linear-time heuristic algorithms like the Landau-Vishkin algorithm
(LV)~\cite{Landau1986} (in addition to traditional algorithms like SW and NW) can be applied to the
sequence alignment problem in SNAP~\cite{Zaharia11}, to compute edit distance accurately up to a
particular number of mismatches (assuming that correct alignments have lower numbers of mismatches).
\cref{algo:basic_alignment_skeleton} describes the skeleton of these heuristic accelerated
algorithms for single-ended read alignment~\cite{PairEndWebLink}. The definitions of the
\texttt{Build\_Index}, \texttt{Candidate\_Locations}, \texttt{Edit\_Distance}, and
\texttt{Find\_Config} functions define different variants of these algorithms. For example,
\cref{tab:alignment_pieces} defines the BWA-MEM and SNAP alignment tools by substituting these
placeholder functions with specific algorithms.

We performed a profiling study of the SNAP aligner on an in-sillico (from an Illumina HiSeq 2500)
whole human genome~\cite{Stephens16} with $50\times$ coverage (i.e., each nucleotide of the
reference is backed by an average of 50 reads that align to that base) on the Blue
Waters\cite{BWWebLink} supercomputer. We chose the SNAP aligner in particular because it is
significantly faster than other alignment tools like BWA and Bowtie. Also, as the LV algorithm used
in SNAP has a linear time complexity, its comparison to ASAP as the CPU baseline is much more
challenging. \cref{tab:snap_callgraph} describes the distribution of runtime across for the SNAP
aligner for corresponding steps of \cref{algo:basic_alignment_skeleton}.\footnote{Note that some
steps of the SNAP aligner implementation includes a variety of other miscellaneous tasks, e.g.,
memory allocation, IO. These are collectively described in the ``Misc'' category. Also note, the
SNAP aligner is optimized to perform asynchronous pre-fetch based disk IO. Hence wait time for IO is
minimized.} These measurements, along with static analysis of \cref{algo:basic_alignment_skeleton},
show the following:
\begin{enumerate}[noitemsep,nolistsep,leftmargin=*]
  \item The LD computation corresponds to nearly $60\%$ of the running time of the SNAP aligner.
  \item The LD computation is one of the most frequently called algorithmic kernels in
  the alignment process (on average called $54.1$ times per read).
  \item The LD kernel is used to build a total ordering of all candidate locations for a read in
  the reference; refer to Line~\ref{algo:line:edit_distance} of 
  \cref{algo:basic_alignment_skeleton}.
  \item The backtrack-based alignment~\cite{Smith81, Needleman1970} is computed only for the 
  best-matched location in the reference.
  \item The remaining portion of SNAP's runtime (after the LD computation) is spent in either memory
  or IO bound computation (e.g., hash table look-ups and reading/writing files). This part is 
  unsuitable for acceleration on PCIe-based devices because of the time-cost associated with 
  performing data transfer over the bus.
\end{enumerate}

\begin{algorithm}[!t]
    \caption{Algorithmic skeleton for single-ended short-read-alignment algorithms.}
    \label{algo:basic_alignment_skeleton}
    \small
    \SetAlgoLined
    \LinesNumbered
    \KwData{NGS Read Dataset, Reference Genome}
    \KwResult{Aligned positions and mapping of reads in Reference Genome}
    $ngsdata \leftarrow$ Set of reads\;
    $reference \leftarrow$ String(s) corresponding to a reference\;
    $index \leftarrow$ \texttt{Build\_Index}$(reference)$\;
    $alignment \leftarrow \emptyset$\;
    \For{$read \in ngsdata$} { \label{algo:line:loop}
        $locs \leftarrow$ \texttt{Candidate\_Locations}$(read, index)$\; 
        \label{algo:line:candidate_location}
        $opt \leftarrow \argmin_{loc \in locs} ($\texttt{Edit\_Distance}$(read, 
        loc))$\; \label{algo:line:edit_distance}
        $config \leftarrow$ \texttt{Find\_Config}$(read, opt)$\; \label{algo:line:config}
        $alignment \leftarrow alignment~\cup~config$\;
    }
    \Return $alignment$\;
\end{algorithm}

\begin{table}[!bp]
    \centering
    \vspace{-10pt}
    \caption{Distribution of runtime across the steps of 
        \cref{algo:basic_alignment_skeleton} for the SNAP tool aligning an in-sillico 
        human 
        genome with $50\times$ coverage.}
    \begin{tabular}{lll}
        \toprule
        \textbf{Lines in \cref{algo:basic_alignment_skeleton}} & \textbf{\% of runtime} & 
        \textbf{\# of calls} \\
        \midrule
        Line~\ref{algo:line:loop} & 6.79 & $1.5\times 10^{10}$\\
        Line~\ref{algo:line:candidate_location} & 18.59 & $6\times 10^{10}$\\
        Line~\ref{algo:line:edit_distance} & 59.22 & $8.3\times 10^{11}$\\
        Line~\ref{algo:line:config} & 9.25 & $1\times 10^{10}$\\
        Misc & 6.15 & --\\
        \bottomrule
    \end{tabular}
    \label{tab:snap_callgraph}
\end{table}
    
    \section{Design of the ASAP Accelerator} \label{sec:asap}

This section describes the approximation algorithm that drives the design of ASAP, provides a proof
for its correctness, and describes its implementation in programmable hardware.
\cref{sec:asap_delay_idea} briefly summarizes the RaceLogic paper~\cite{Madhavan2014}, describing an
formalizing the encoding the computation of LD scores into circuit propagation delay.
\cref{sec:asap_approximation} describes the approximation at the heart of ASAP: using the ability to
directly tune the performance of the algorithm to input-data characteristics (i.e., using circuit
propagation delays encode both the algorithm and its computation time), we show a method to chose
appropriate propagation delays to compute approximate answers for LD while maintaining their total
ordering (i.e., satisfy the application invariant for correctness). Finally,
\cref{sec:asap_fpga,sec:asap_capi} describes the ASAP FPGA implementation.

\subsection{Encoding LD Computation in Circuit Propagation Delays} \label{sec:asap_delay_idea}
\begin{figure}[!t]
    \centering
    \includegraphics[width=\columnwidth]{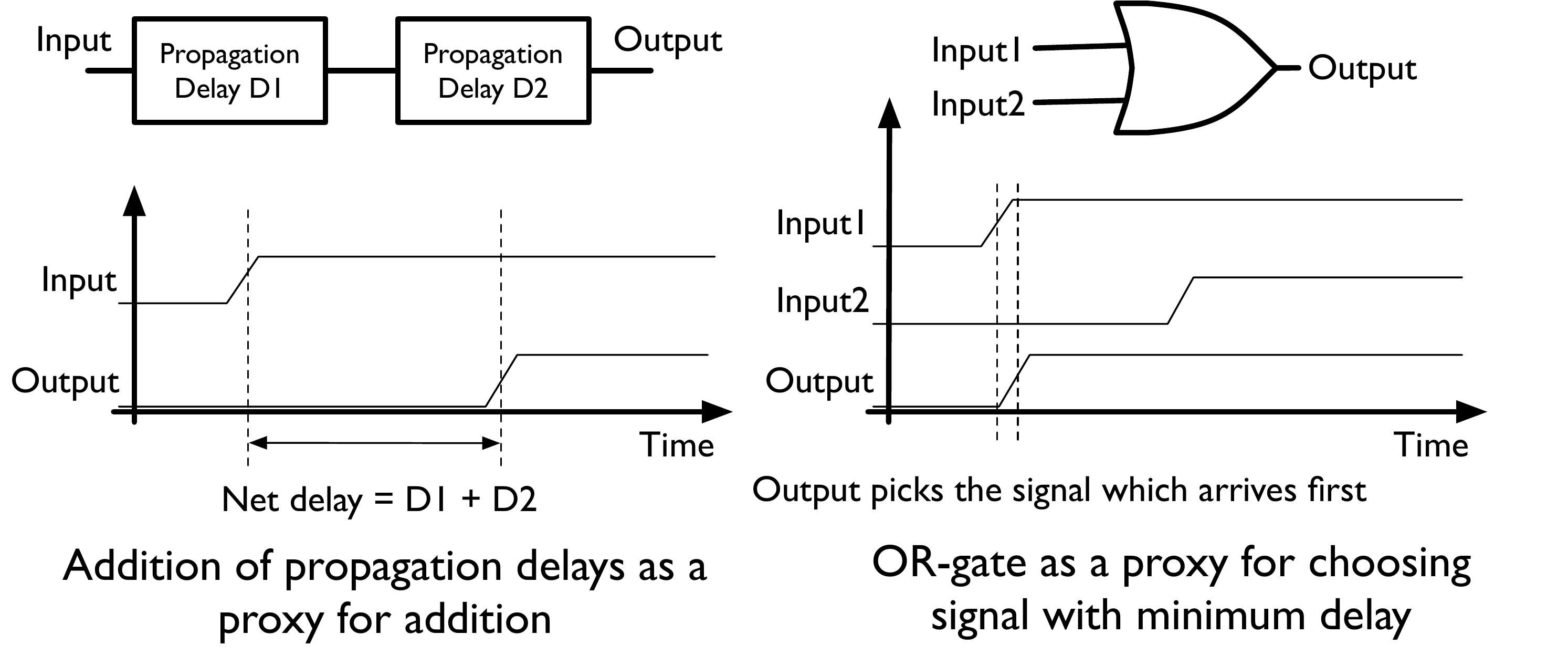}
    \caption{Computing with propagation delays: Delay-based proxy for the addition operator is 
        a series connection, and the proxy for the $\min$ operator is the \texttt{OR} gate.}
    \label{fig:propagation_delays}
\end{figure}

\begin{figure}[!bp]
    \centering
    \includegraphics[width=\columnwidth]{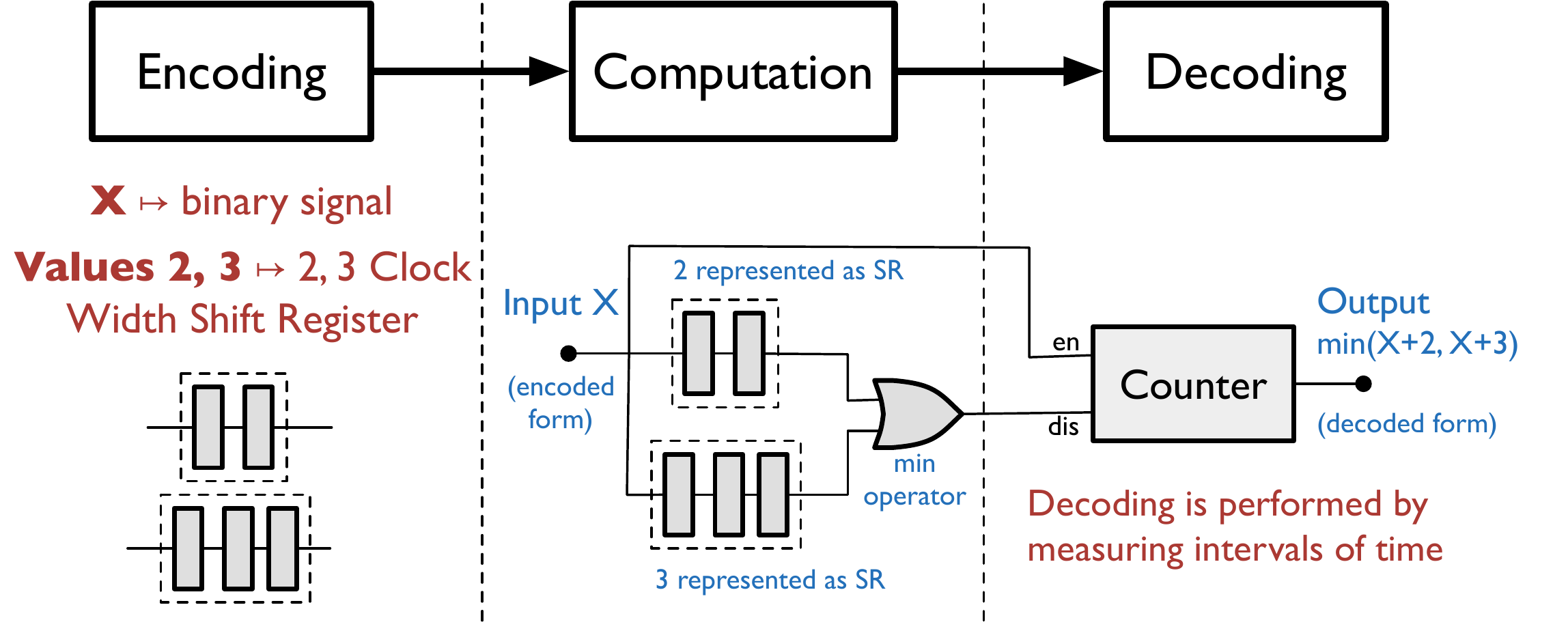}
    \caption{Example of the encoding, computation, and decoding phase for computing ``$\min(X +
        2, X + 3)$'' using the circuit-delay proposed in RaceLogic~\cite{Madhavan2014}. Note that we
        present this example using shift-registers for delay elements as opposed to comparators
        proposed in~\cite{Madhavan2014}.}
    \label{fig:rl_example}
\end{figure}

The core idea is to map addition and minimization, the two mathematical operators necessary for the
recursive computation defined in \cref{eqn:basic_levenshtein_distance}, to particular topologies of
circuit elements. \cref{fig:propagation_delays} illustrates the mapping explained here:
\begin{enumerate}[noitemsep,nolistsep,leftmargin=*]
    \item If circuit elements are combined in series, the net propagation delay of a signal is the
    sum of the propagation delays for all of the individual elements. This construction is a proxy
    for addition.
    \item If two circuit elements are connected to an OR gate, the signal that
    emerges out of the OR gate corresponds to the signal that arrived first at the gate.
    This construction is a proxy for the minimization operator (in particular, the rising edge of the OR gate's output computes a minimization in time).
\end{enumerate}
For example, \cref{fig:rl_example} demonstrates the computation of ``$\min(X + 2, X + 3)$'' using
the aforementioned delay based computing. In the example, $X$ corresponds to an arbitrary input
signal that is represented in the delay encoding, the 2- and 3-length shift register serving as the
delay element implementing the $\boldsymbol{\cdot} + 2$ and $\boldsymbol{\cdot} + 3 $ operator
respectively, the OR gate serves as the minimization operator and the counter serving as the
decoder.

We formalize this delay based computation succinctly in the following lemma.
\begin{lemma}
 Propagation-delay-based computation can occur on a tropical semiring structure
 $\mathcal{T}$ over $\{0\} \cup \mathbb{Z}^{+}$ (i.e., time measured in clock ticks) that 
 defines a binary addition operation, a minimization operator (using an OR gate), and a 
 maximization operator (using an AND gate).
\end{lemma}

\begin{figure}[!t]
    \centering
    \includegraphics[width=\columnwidth]{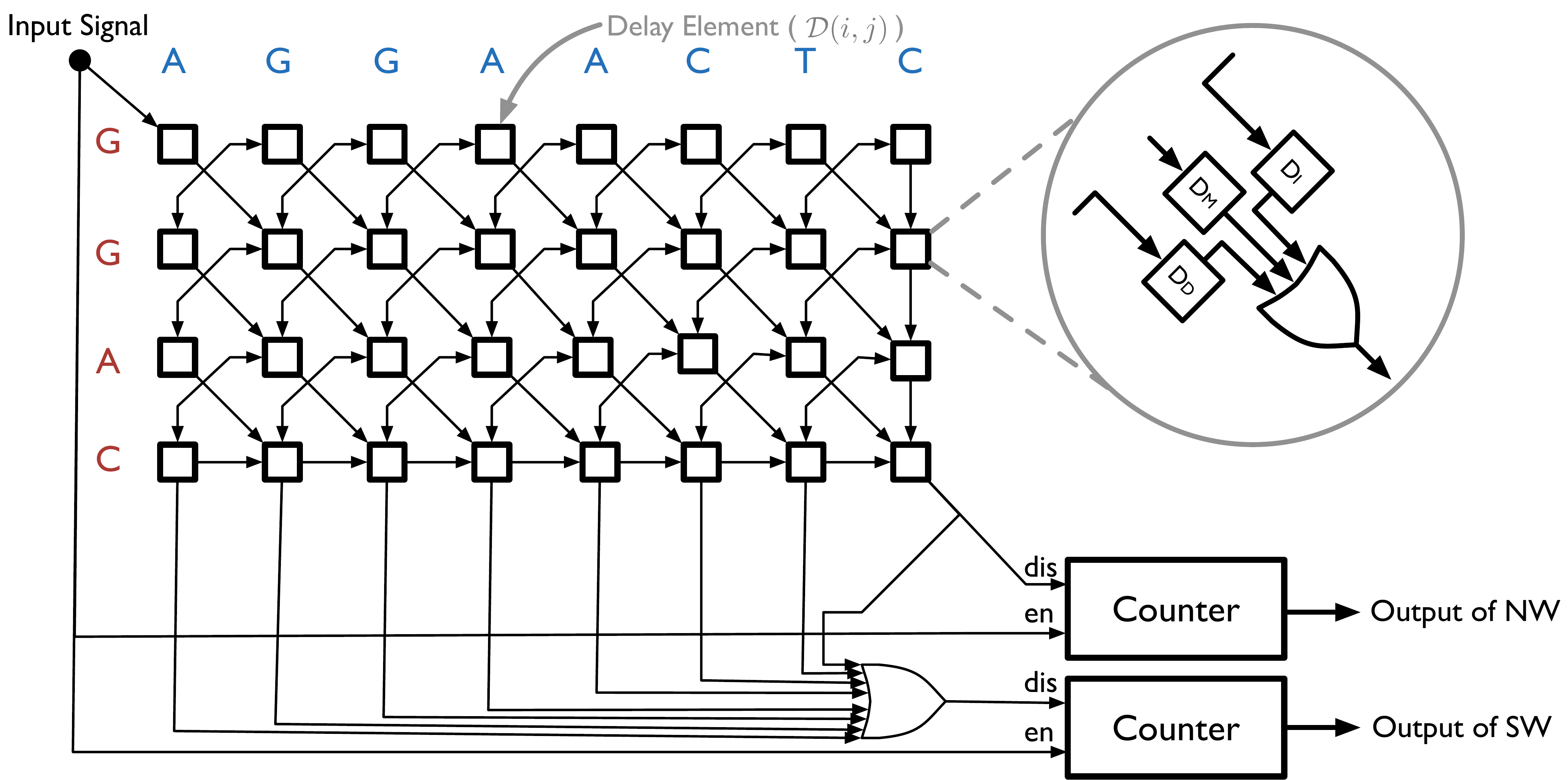}
    \caption{High-level design of the ASAP accelerator to compute the minimum edit distance 
        between two strings. The accelerator lattice is of size $l_Q \times l_R$, where $l_Q$ and 
        $l_R$ are the sizes of the query and reference, respectively.}
    \label{fig:asap_array}
\end{figure}

The delay-based proxies for the addition and minimization operators can be used by replacing the LD
values $S(i,j)$ in \cref{eqn:basic_levenshtein_distance} with the equivalent propagation delays. The
resulting circuit represents the application of the addition and minimization operators in the
computation of $S(i,j)$. \cref{fig:asap_array} shows the structure of the circuit that produces this
computation.  It is composed of a lattice of $l_Q \times l_R$ delay elements (DEs). The connections
in the lattice build on the recursive definition of $S$: each DE $\mathcal{D}(i, j)$'s inputs are
connected to the outputs of the preceding elements $\mathcal{D}(i-1, j-1)$, $\mathcal{D}(i-1, j)$,
and $\mathcal{D}(i, j-1)$, and its outputs are connected to the input of $\mathcal{D}(i+1, j+1)$,
$\mathcal{D}(i+1, j)$, and $\mathcal{D}(i, j+1)$. At a high level, each DE is composed of three
delay blocks:
\begin{enumerate*}
    \item $D_M$ (delay due to match or mismatch at $(i,j)$),
    \item $D_I$ (delay due to insertion at $(i,j)$), and
    \item $D_D$ (delay due to deletion at $(i,j)$).
\end{enumerate*}
This design is specialized for FPGAs in \cref{sec:asap_fpga})
    
The computation can be started by injecting a high signal (logic value $1$) at the inputs of index
$\mathcal{D}(0, 0)$ in the array. The time-encoded value of the LD is then found by measuring the
propagation delay of the signal exiting the array of delay elements. Note that the delay-based
computation can be applied to all variants (SW, NW, and LV) of the LD computation as follows.
\begin{enumerate}[noitemsep,nolistsep,leftmargin=*]
	\item The delay-based version of the \emph{SW} variant can be computed by measuring the delay
	between the introduction of the input signal in the lattice, and its emergence at any of the
	delay elements on the last row, \emph{i.e.,} $(l_R, -)^{th}$ DE. \cref{fig:asap_array} 
	illustrates this configuration.
	\item The delay-based version of the \emph{NW} variant can be computed by measuring the delay
	between the introduction of the input signal in the lattice, and its emergence at the
	$(l_R, l_Q)^{th}$ DE. This configuration is also shown in \cref{fig:asap_array}.
	\item The delay-based version of the \emph{LV} variant can be computed by assigning the maximum
    permissible LD as the result of the computation. This represents the ``timeout'' with which the
    signal wavefront will emerge from the DE lattice. If the timeout is triggered, the maximum 
    value of LD, as set by the user, is used as the result of the computation. One delay element
    and one AND gate (not shown in the \cref{fig:asap_array}) suffice to implement the 
    timeout.
\end{enumerate}

\subsection{Approximating LD Computations in ASAP}\label{sec:asap_approximation}

A key aspect of the aforementioned method is the mapping of gap-penalty parameters
(\mbox{$\Delta$-parameters}) to their corresponding circuit delays. The ASAP accelerator uses this
mapping both to encode the approximation (mentioned in \cref{sec:introduction}), and to reduce the
time taken to do the ``match''-based computation. Both actions are formally stated below.

\begin{definition} \label{def:delay_encoding}
     A \emph{Delay Encoding Function} $\mathcal{E}:\mathbb{R} \rightarrow \mathcal{T}$ is a mapping
    between the set of real numbers and its propagation-delay-based representation. $\mathcal{E}$ is
    constrained to obey the Cauchy functional equation ($\mathcal{E}(x+y) = \mathcal{E}(x) +
    \mathcal{E}(y)$).
\end{definition}

More general delay encoding functions can be considered, for example in analog circuits, where
circuit elements do not exhibit linear behavior for all inputs. We constrain ourselves to those that
satisfy the Cauchy functional equation (CFE) because of simplicity in proving of correctness of the
transformation. Although the domain of $\mathcal{E}$ can be the set of real numbers $\mathbb{R}$,
the ASAP implementation presented in this paper uses integer or rational gap penalties which can be
easily mapped to integer delay values (which can further be represented as a multiples of the clock
width).

\begin{definition} \label{def:delta_params}
    A $\delta$-parameter is the time-encoded representation of a user-inputted $\Delta$-parameter. 
    That is
    \begin{eqnarray}
     \delta(\text{Insert}) &=& \mathcal{E}(\Delta(\text{Insert})) \nonumber \\
     \delta(\text{Delete}) &=& \mathcal{E}(\Delta(\text{Delete})) \nonumber \\
     \delta(\text{Match}) &=& \mathcal{E}(\Delta(\text{Match})) \nonumber \\
     \delta(\text{Mismatch}) &=& \mathcal{E}(\Delta(\text{Mismatch}))
     \label{eqn:encoding_mapping}
    \end{eqnarray}
    These parameters are used to define the delays in the $D_M$, $D_I$, and $D_D$ blocks. Note that
    we have assumed that $\Delta(\text{Match})=0$, and thus $\delta(Match)=\mathcal{E}(0)$ is also~0
    based on \cref{def:delay_encoding}.
\end{definition}

\begin{figure*}[t]
    \centering
    \includegraphics[width=\textwidth]{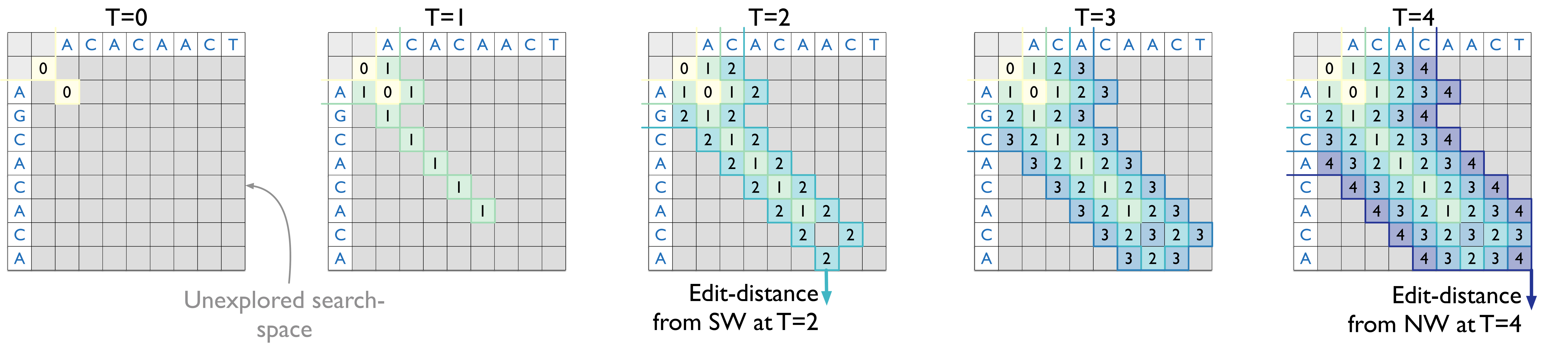}
    \caption{An example of the ASAP accelerator processing the same inputs used in 
    \cref{fig:dynamic_prog_matrix}. The signal wavefront is shown progressing through the 
    ASAP lattice until the outputs of the SW and NW algorithms are produced in 2 and 4 clock 
    cycles, respectively. The values in the matrix represent the clock cycles in which the 
    corresponding DEs were enabled.}
    \label{fig:wave_propagation_example}
    \vspace{-10pt}
\end{figure*}

Based on definitions~\ref{def:delay_encoding} and~\ref{def:delta_params}, we now show that any
encoding of $\delta$-parameters based on $\mathcal{E}$ produces the same ordering of LDs as the
original algorithm.

\begin{lemma} \label{lem:asap_equivalence}
    When a query string $Q$ and a reference string $R$ are compared under the traditional
    (see \cref{eqn:basic_levenshtein_distance}) and delay-based algorithm for computing LD at
    loci $l_1, \dots, l_n$ of the reference, to produce LDs $e_1,\dots, e_n$ and propagation delays
    $d_1, \dots, d_n$, respectively, then $d_i = \mathcal{E}(e_i)$, and consequently
     \[
         e_i \leq e_j \iff \mathcal{E}(e_i) \leq \mathcal{E}(e_j) \iff d_i \leq d_j~\forall i,j.
     \]
\end{lemma}

\cref{lem:asap_equivalence} is sufficient to show that using the ASAP accelerator to compute LD in
the context of \cref{algo:basic_alignment_skeleton} (in line~\ref{algo:line:edit_distance}; i.e.,
using an ``$\argmin$'' operator over the results of multiple executions of the ASAP accelerator)
produces the same result as the traditional algorithm (without requiring the computation of the
inverse for $\mathcal{E}$). A key observation in the formalism of $\mathcal{E}$ is that the choice
of the numerical values of $\delta$ can be tuned to directly change the performance of the
accelerator, as they corresponds to circuit propagation delays. That is, the parameters and inputs
to the accelerator jointly define the net propagation delay of the circuit. Below we demonstrate one
such transformation, which forms the core of the approximation used in ASAP.

\begin{lemma} \label{lem:asap_tuning}
     When a query string $Q$ and a reference string $R$ are compared at loci $l_1, \dots, l_n$ of
    the reference, they produce LDs $e_1,\dots, e_n$ for gap penalties $\Delta$, and LDs $e_1',
    \dots , e_n'$ for gap penalties $\Delta + k$, for some number $k$. The $e_i'$ obey the
    relationship: $e_i' = e_i + n_ik$, for some $n_i \in \mathbb{Z}$ such that ($n_i \geq 0) \land
    (e_i \leq e_j \iff n_i \leq n_j)$, and consequently
     \[
         e_i \leq e_j \iff e_i' \leq e_j'~\forall i,j.
     \]
\end{lemma}

Our algorithm for the approximation at the core of ASAP uses Lemmas~\ref{lem:asap_equivalence}
and~\ref{lem:asap_tuning} to select values of the delay-encoded parameters that correspond to
minimizing the time taken to process a dataset. For example, to optimize performance for our
observed case of most nucleotides corresponding to ``matches,'' we modify the gap-penalties to set
the match penalty (i.e., $\delta(\text{Match})$) to 0 cycles\footnote{True ``0 cycle'' propagation
delay is not possible because of finite combinational and wire delays in the circuit. Here we imply
that the computation is done in combinational logic, whose propagation delay is much much lower than
the clock width of the circuit (i.e., 0 time). This is explained further in \cref{sec:asap_fpga}.}.
This transformation uses a two-step process to convert (encode) user-inputted $\Delta$-parameters
into $\delta$-parameters:
\begin{enumerate}[noitemsep,nolistsep,leftmargin=*]
    \item $\Delta \mapsto \Delta + k$, choosing $k$ so that $\Delta(\text{Match}) = 0$ after the 
    transformation;
    \item $\Delta + k \mapsto \mathcal{E}(\Delta + k)$, with $\mathcal{E}(x) = mx$ to produce the 
    required delay value.\footnote{The choice of $k$ and $m$ has to ensure that none of the encoded
    gap penalties are negative. As the encoded values represent circuit propagation delays, negative
    numbers are meaningless.}
\end{enumerate}
As a result, the parameters in the LD algorithm are tweaked to better suit the delay-based
computation hardware. The answer (i.e., the exact values of LD) produced by this approximate version
of the algorithm is not identical to that produced by the original algorithm. However, based on the
aforementioned lemmas, we can see that the total ordering created by the approximated LDs is
identical to that of the original algorithm. Furthermore, assuming that most nucleotide comparisons
are matches (which is true for the indexed reference-based techniques described in
\cref{sec:ld_and_alignment}), this encoding ensures that (almost) zero time is taken to explore
large portions of the search space that correspond to matches. We explore the relation of this
optimization to timing closure on the FPGA design in \cref{sec:asap_fpga}. In other re-sequencing
experiments, where ``matches'' do not represent the common computation, a user can set $\delta(k) =
0$ for $k \in \{\text{Insert, Delete, Mismatch}\}$. Note that in our formulation of the problem (as
described in \cref{sec:ld_and_alignment}), $\Delta(\text{Match})$ is required to be the minimum
positive value amongst all the $\Delta$-parameters.

\begin{figure*}[!t]
    \centering
    \includegraphics[width=\textwidth]{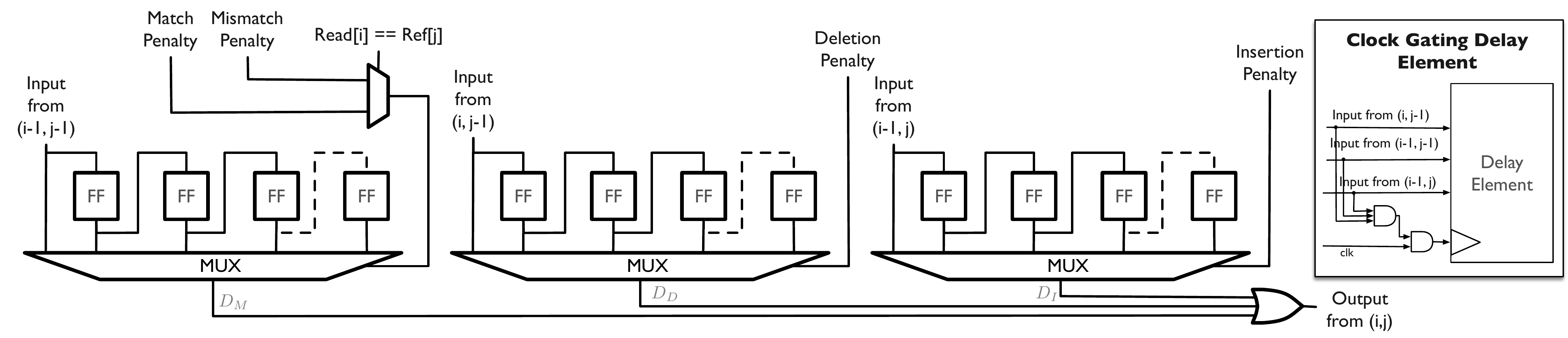}
    \caption{Design of a single delay element $\mathcal{D}$ in ASAP. The DE is composed of three 
        separate delay units corresponding to $D_M$, $D_I$, and $D_D$ in 
        \cref{fig:asap_array}.}
    \label{fig:delay_element}
    \vspace{-10pt}
\end{figure*}

Consider the example of computing the LD between the strings \texttt{AGCACACA} and
\texttt{ACAACAACT}, presented in \cref{sec:ld_and_alignment}. Based on our encoding mechanism ($k=0,
m=1$), we compute the $\delta$-parameters of the ASAP~accelerator as $\delta(\text{Match}) = 0$,
$\delta(\text{Mismatch}) = 2$, and $\delta(\text{Insert}) = \delta(\text{Delete}) = 1$.
\cref{fig:wave_propagation_example} illustrates the propagation of the signal wavefront through the
ASAP accelerator for that example. The accelerator produces an output for the SW notion of LD (local
alignment) in two clock cycles and the NW notion of LD (global alignment) in four clock cycles. The
figure shows the portion of the array explored and the value of the propagation delay at each
element $\mathcal{D}(i, j)$ of the lattice. Note that some portions of the array are not explored at
all (e.g., for SW and NW, only $25$ and $53$ DEs out of a total of $81$ are triggered,
respectively). This design thus provides a large savings in both time (using ``zero delay'' circuit
components for the most commonly used computation) and power (clock-gating unused DEs with their
input signals ensures minimal power usage) compared to traditional methods.

To summarize, using the encoding of $\delta$-parameters described in this section, the ASAP 
accelerator has two clear advantages over traditional techniques:
\begin{enumerate}[noitemsep,nolistsep,leftmargin=*]
    \item \emph{Faster Processing}: One can explore large portions of the search space in a small 
    amount of time by setting delay parameters appropriately.
    \item \emph{Energy Savings}: DEs in the ASAP lattice are used only when their output can 
    contribute to the answer; otherwise, they are switched off to save energy. This can be 
    accomplished by clock-gating the DEs with their input signal.
\end{enumerate}

\subsection{ASAP: The FPGA Implementation} \label{sec:asap_fpga}

\subsubsection{Why FPGA?}
The techniques discussed so far in the paper represent an approximation technique and architecture,
one which can be implemented ASICs, FPGAs, or any other platform. The original RaceLogic design was
demonstrated in simulation as an ASIC~\cite{Madhavan2014}. However, some key characteristics of the
short-read alignment problem and the ASAP architecture make ASAP particularly suitable for FPGAs, as
they offer programmability and reconfiguration. The ASAP accelerator is runtime-programmable only
for changing the values of gap penalties. The input data size, which defines the size of the
accelerator lattice, is fixed at compile time. To allow users to sweep experiment such
``meta-parameters'' (i.e., input data size, gap-penalty bit-width, and input encoding), ASAP is
designed to be re-synthesized and re-programmed on an FPGA. Potentially, the use of partial
reconfiguration can allow users to change these parameters on the fly. We leave this possibility for
future work. We discuss the advantages of the ASAP design compared to the commonly used systolic
array based design (e.g.,~\cite{Lipton1985, Hoang1993, Guccione2002, Zhang2007, Ahmed2015}) in
\cref{sec:related_work}.

\subsubsection{Design of a Delay Element}
The overall architecture of the ASAP accelerator is shown in \cref{fig:asap_array}.
\cref{fig:delay_element} shows the design of a single DE. A DE utilizes sequential
logic in the form of a shift-register to add a user-specified amount of delay. Each DE has
\begin{enumerate*}
    \item three input signals (representing input wavefront) that connect it to its preceding DEs 
    in the grid,
    \item two input signals representing the nucleotides being compared by the element, and
    \item three input signals representing the $\delta$-parameters.
\end{enumerate*}
Each DE has one output signal representing the propagated wavefront after the delay has been added.
The match, mismatch, insertion, and deletion delay penalties are defined in terms of multiples of
the clock period. When the input signal wavefront first reaches an element, it is propagated through
a shift register to create delay. Based on the gap penalty specified for match/mismatch, insertion
and deletion, the DE propagates the input signals to the output.  The output of each flip-flop in
the shift register is muxed to allow for the selection of the bit corresponding to the gap-penalty
of the block (illustrated in \cref{fig:delay_element}). The ASAP array allows the user to program
(i.e., dynamically set at runtime) the values of the select lines of these \texttt{MUX}s. This
provides the ASAP array with a degree of programmability, allowing it to be reused across
computations that merely require re-parameterization of the gap-penalties. Changes in input-sizes,
or the dynamic range of the gap penalties (i.e., number of bits required to represent the
gap-penalties) requires a re-synthesis and reconfiguration of the accelerator on the FPGA.

As described in the motivating example for the ASAP accelerator given in
\cref{fig:wave_propagation_example}, the power of the ASAP accelerator is that it can explore a
large portion of the search space of possible mappings between the query string and the reference
within a clock cycle by setting $\delta(\text{Match}) = 0$. This improvement in computational speed
can be coupled with a decrease in energy consumed by the accelerator by clock-gating the DE 
(illustrated in \cref{fig:delay_element}) with the input signal.

\begin{figure}[!t]
    \centering
    \includegraphics[width=\columnwidth]{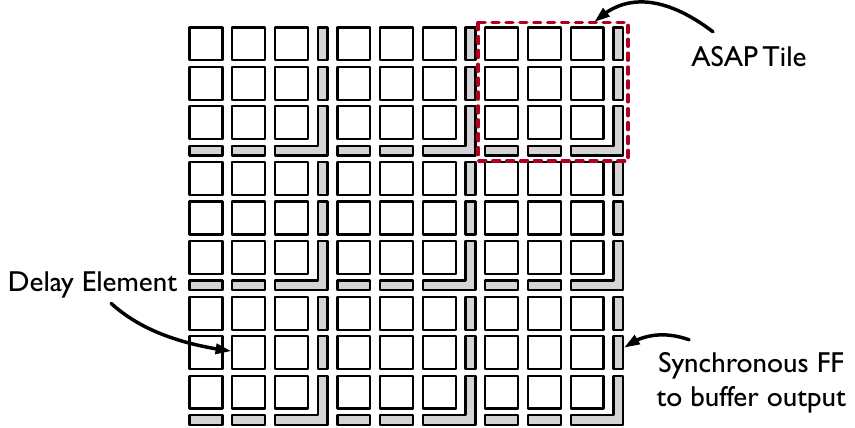}
    \caption{The architecture of the ASAP accelerator in terms of tiles whose output is 
        buffered by clock synchronous flip-flops (FFs).}
    \label{fig:asap_flip_flops}
\end{figure}

The approach mentioned above has problems with long chains of combinational logic and may lead to
timing violations on large lattices of DEs. To get around this problem, larger lattices of delay
elements are composed by using the smaller tiles of ASAP accelerators (for which the timing
violations do not occur) and by adding a sets of clock-triggered flip-flops between the tiles to
break the chains of combinational logic (see \cref{fig:asap_flip_flops}). Further, the diagonal tile
crossing (i.e., the flip-flops at the lower right corner of the tile) corresponds to a 2 cycle delay
(i.e., two flip-flops in serial). Although the additions of the tile flip-flops changes the results
of ASAP from what was described in the last section, the overall total-ordering is preserved, as
this constitutes a constant addition of delay to all outputs of the ASAP accelerator. Each tile is
synthesized, optimized, and placed-and-routed separately by defining separate design partitions.
This approach prevents the compiler from performing optimizations across partition
boundaries~\cite{XilinxParitions}. This approach also ensures that unintended wiring delays do not
creep into the netlist of the ASAP accelerator.

The counter that decodes the delayed signal output from the ASAP lattice (shown in
\cref{fig:asap_array}) is designed based on a computation of the number of clock cycles for the
signal wavefront to emerge from the lattice. The bit-width of this counter, $N_o$, is calculated
from the sizes of the input strings and the user-input gap-penalty parameters, and is given by
\[
 N_o = \left\lceil \log_2 \min\left\{
     \begin{array}{ll}
       \delta_Il_Q + \delta_Dl_R,\\
       \delta_Ml_Q + \delta_D(l_R - l_Q)
     \end{array}
   \right\} \right\rceil.
\]
This expression is an upper bound (albeit a loose one) on the maximum delay caused by a DE. 
    
\subsubsection{Scalability Issues in the ASAP Accelerator}
\begin{figure}[!t]
    \centering
    \includegraphics[width=\columnwidth]{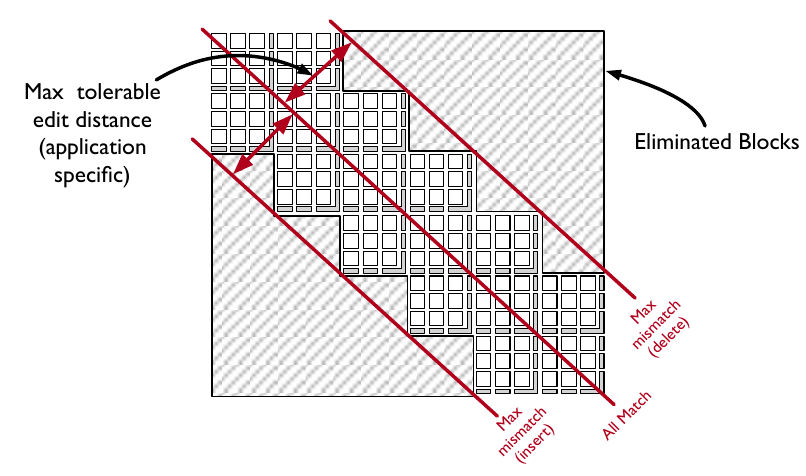}
    \caption{Elimination of unused tiles from the ASAP lattice in the case of LV variant of the 
        LD algorithm}
    \label{fig:asap_removed}
\end{figure}

There are challenges involved in scaling the ASAP 
accelerator to large input sizes and large gap penalties. Those challenges can be addressed as follows:
\begin{enumerate}[noitemsep,nolistsep,leftmargin=*]
    \item \emph{Large Input Sizes.} The size of the reference and read strings being compared in the
    ASAP accelerator plays a role in the size of the lattice defined by the ASAP accelerator. The
    size of the accelerator grows as $O(l_Q \times s)$  with the input size\footnote{This
    corresponds to quadratic growth in size of the ASAP lattice (i.e., $O(n^2)$) when $l_Q = s =
    n$.}. The tile size parameter defines a tunable knob to control the critical combinational path
    in the circuit. It can be used to trade off performance against meeting timing closure as the
    size of the	accelerator grows to a significant portion of the resources available on the FPGA.
    \cref{sec:evaluation} demonstrates our scaling experiments with the accelerator.
    \item \emph{Large Gap Penalties.} A large dynamic range of the gap-penalty values negatively
    affects the ASAP accelerator, as it increases the size of the shift-registers and multiplexers
    in the DE (see \cref{fig:delay_element}). We work around this problem by using BRAM-based
    shift registers, which can be $\sim 10^3$ bits long (without intermediate routing). In general,
    we do not expect large gap penalties to be a problem for genomic sequences (as opposed to
    protein sequences), for which the dynamic range in gap-penalties is low.
    \item \emph{Potentially Unused Tiles.} \cref{fig:wave_propagation_example} shows that a 
    large part of the ASAP array is not involved in computation when the input strings have low LD 
    (which is indeed the case in the short read alignment problem). There are several ways to 
    tackle the problem of unused tiles across the three variants of the LD computation (i.e., SW, 
    NW, and LD). As mentioned earlier, in the case of SW or NW, clock-gating individual delay 
    elements ensures minimal power consumption. Further, in the LV case, as a the worst case LD is 
    specified, we can use this information at compile (in this case synthesis) time to eliminate 
    part of the ASAP lattice that will not contribute to an answer. \cref{fig:asap_removed} 
    illustrates such an elimination on an $18\times 18$ lattice with a maximum of $6$     
    insertions or deletions permitted, resulting in a $56\%$($=\nicefrac{20}{36}\times 100$) 
    reduction in area.
\end{enumerate}

\begin{figure*}[!t]
    \centering
    \includegraphics[width=\textwidth]{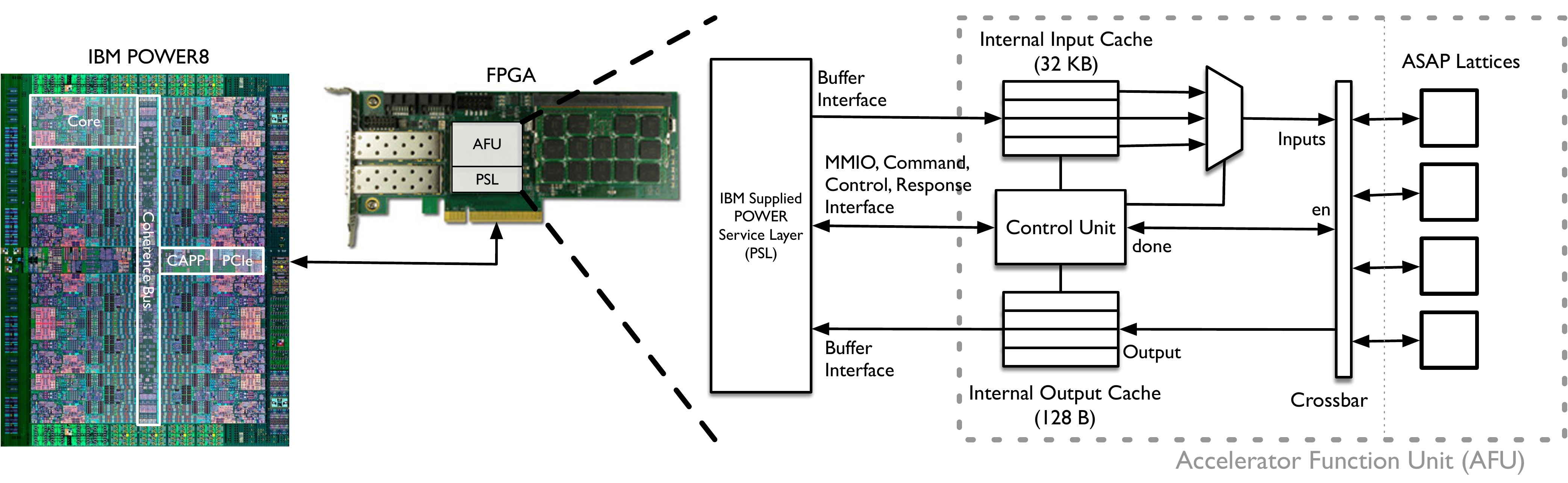}
    \caption{The design of the interface between the host Power8 processor and the FPGA running the 
        ASAP accelerator using the CAPI interface. The diagram assumes an ASAP accelerator that 
        computes on input strings that are 64 nucleotides long and encoded as 2 bits per 
        nucleotide.}
    \label{fig:asap_capi_interface}
    \vspace{-10pt}
\end{figure*}  

\subsubsection{Issues with Timing Closure}
Computing with propagation delays is disadvantaged by the fact that thermal dissipation and
temperature variations at different parts of the FPGA chip to change the physical time associated
with unit delay.  However, the ASAP accelerator is resilient to these thermal changes up to the
maximum operating temperature of the FPGA (i.e., timing violations do not occur). Further, only
delays that are multiples of the clock period can affect the computed LD. The tile length serves as
a tunable knob between runtime performance and worst case negative slack for the circuit. This slack
is enforced by the compiler (e.g., Xilinx Vivado, Altera Quartus) as only values of tile length for
which timing closure can be met can be used in the FPGA. Furthermore, the counters in
\cref{fig:asap_array} that measure edit distance are synchronously triggered by the clock, thereby
ensuring that all delay-based LDs are computed as multiples of the clock cycle.
	
\subsubsection{Encoding Input Sequences}
The implementation of the ASAP accelerator assumed use for genomic data, implying that the entire
alphabet can be represented in two bits (i.e., \texttt{A}, \texttt{C}, \texttt{G} and \texttt{T}).
The bases \texttt{N}, \texttt{-}, \texttt{R}, \texttt{Y}, \texttt{K}, \texttt{M}, \texttt{S}, and
\texttt{W} (which represent an unknown or ambiguous nucleotide) are removed from the alphabet. Our
design could potentially be extended to larger alphabets, e.g., for protein sequence alignment.

\subsection{Host-to-Accelerator Communication via CAPI}  \label{sec:asap_capi}
Communication between the host and accelerator is implemented using the CAPI
interface~\cite{CAPI_UserManual2014, CAPI2015} provided on an IBM Power8 CPU. The CAPI interface
gives an accelerator (a PCIe-attached FPGA) coherent access to the virtual address space of a
process running on the host CPU, with all address translations from virtual to physical memory done
in the CPU.  \cref{fig:asap_capi_interface} shows the interface and mechanism by which the host CPU
communicates with the ASAP accelerator. The Power8 is a superscalar symmetric multiprocessor, that
has 12 cores per chip, with up to 8 hardware threads per core. All cores have access to shared
memory through a PowerBus (shared memory bus). The Coherent Attached Processor Proxy (CAPP) enables
the interface (CAPI) by maintaining a directory of cache lines held by the processor and providing
coherency by snooping the PowerBus on behalf of the accelerator (or any other PCIe device). The PCIe
host bridge provides connectivity between the CAPP and the Power Service Layer (PSL) on the FPGA
over the PCIe bus. The PSL on the accelerator acts as a proxy for the CAPI protocol on the FPGA,
communicating between the CAPP and the Accelerator Functional Unit (AFU). The AFU contains the
custom acceleration logic and reads/writes coherent data across the PCIe.  The PSL unit runs at the
same speed as the PCIe bus (250 MHz). It contains a memory management unit (MMU) to handle address
translation on the accelerator side on its copy of the processor's cache directory.

The AFU interacts with the PSL to provide word-level read and write commands. If these requests are
made to cache lines (which are 1024 bits long) in a shared or exclusive state on the device, they
are served locally. Otherwise the PSL interacts with the CAPP over the PCIe bus to attempt virtual
to physical address translation, loading of the cache line from main memory (if it is already not
present in the processor's cache), moving (or copying of) the cache line to the PSL, and changing
the coherence of the cache line in the processor's directory~\cite{CAPI_UserManual2014,
Jaspers2015}.  We use the AFU in dedicated mode, meaning only one MMU context is supported by the
accelerator.  That is, only one user-space process can use the accelerator at one time.

\cref{fig:asap_capi_interface} shows the configuration of the interface to the PSL for an ASAP
accelerator that computes on two 64-bp strings, with each nucleotide encoded by two bits. Hence the
accelerator takes 256-bit inputs ($64 \text{ bp} \times 2~\text{bits/bp} \times 2$) and produces a
propagation delay measurement encoded in 32 bits (to keep with the signed integer implementation in
short-read aligner), which is the number of clock cycles for the signal to emerge from the ASAP
accelerator (depending on whether the SW or NW algorithm is used). There is an internal 32~kB cache,
which has a 1024-bit input port connected to the PSL, and a 1024-bit output port that is connected
to the input of the ASAP accelerator. This cache is configured in a modified FIFO configuration;
each entry in the FIFO contains multiple input cases (in this case, four). A $4 \times 1$ MUX
controlled by the AFU control unit is responsible for producing 256 bits at a time from the 1024-bit
input. The AFU packs the 32 bit outputs from the ASAP array into 1024 bit cache-lines before writing
them back to the address space of the host over DMA. The AFU uses the work element descriptor
(WED;~\cite{CAPI_UserManual2014}) to communicate the pointer to the input and output, as well as the
progress of the accelerator.
    
    \begin{figure}[!bp]
    \centering
    \includegraphics[width=\columnwidth]{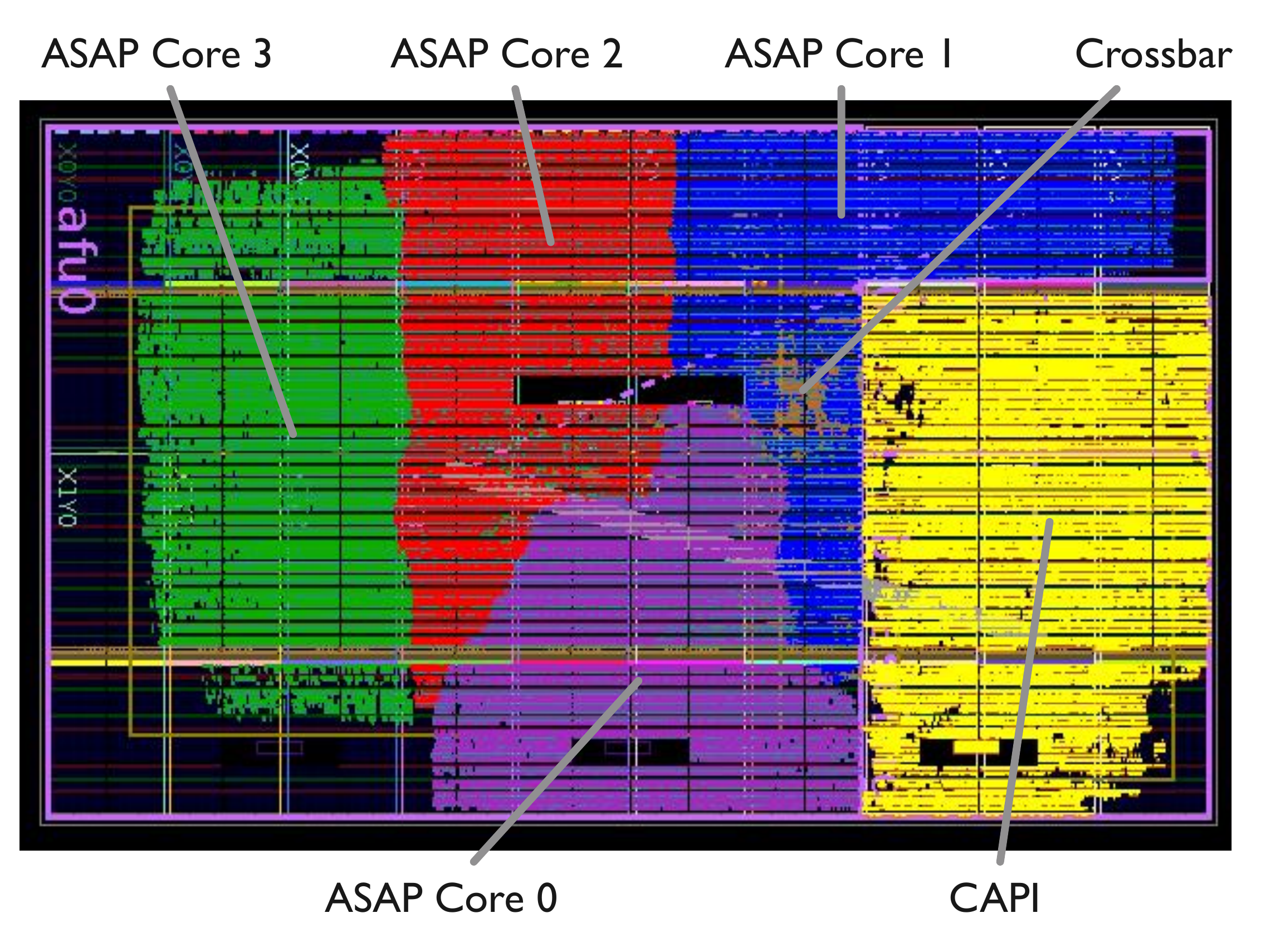}
    \caption{Layout of the accelerator on the Xilinx Virtex 7 XC7VX690T FPGA. The design 
        implemented above has 4 instances of the ASAP accelerator and the IBM CAPI interface for 
        host-accelerator communications.}
    \label{fig:layout}
\end{figure}

\section{Evaluation and Discussion} \label{sec:evaluation}
The ASAP accelerator is implemented in Chisel~\cite{Bachrach2012} and can potentially be compiled
across FPGAs and CAD tools provided by Xilinx and Altera. The host-accelerator interface (which
utilizes IBM CAPI) is implemented in VHDL and is specific to an IBM Power8 S824L system with an
Alpha-Data ADM-PCIE-7V3 board (that uses a Xilinx Virtex 7 XC7VX690T FPGA) clocked at $250$ MHz. All
measurements (baseline CPU as well as FPGA-based) were done on this machine. \cref{fig:layout}
illustrates the layout of four ASAP lattices and the CAPI based interface on the Virtex 7 FPGA
mentioned above.

All inputs for the experiments presented in this section are derived from the human reference genome
\texttt{hg38} by simulating~\cite{Stephens16} 100 million reads of appropriate length. The read
simulation introduced random mutations and simulated sequencing-error models from an Illumina HiSeq
2500 with a $0.1\%$ sequencing error rate. We verified the correctness of our implementation
through comparison with
\begin{enumerate*}
    \item answers generated from the software tools (i.e., in this case SNAP~\cite{Zaharia11});
    \item the ground truth values generated by the simulator.
\end{enumerate*}

The remainder of this section is organized as follows. In \cref{sec:area_based_scaling}, we discuss
the resource consumption in ASAP with varying input sizes. Then, in \cref{sec:asap_perf}, we discuss
a micro-benchmark performance comparison of a single ASAP lattice (configured in SW mode) with a
software-based SW implementation. \cref{sec:capi_perf} discusses the performance implications of the
CAPI interface. Finally in \cref{sec:fpga_resource,sec:snap_perf}, we discuss the power and
performance characteristics (respectively) of an end-to-end application of ASAP (configured in LV
mode) in the SNAP aligner.

\begin{figure}[!t]
    \centering
    \includegraphics[width=0.8\columnwidth]{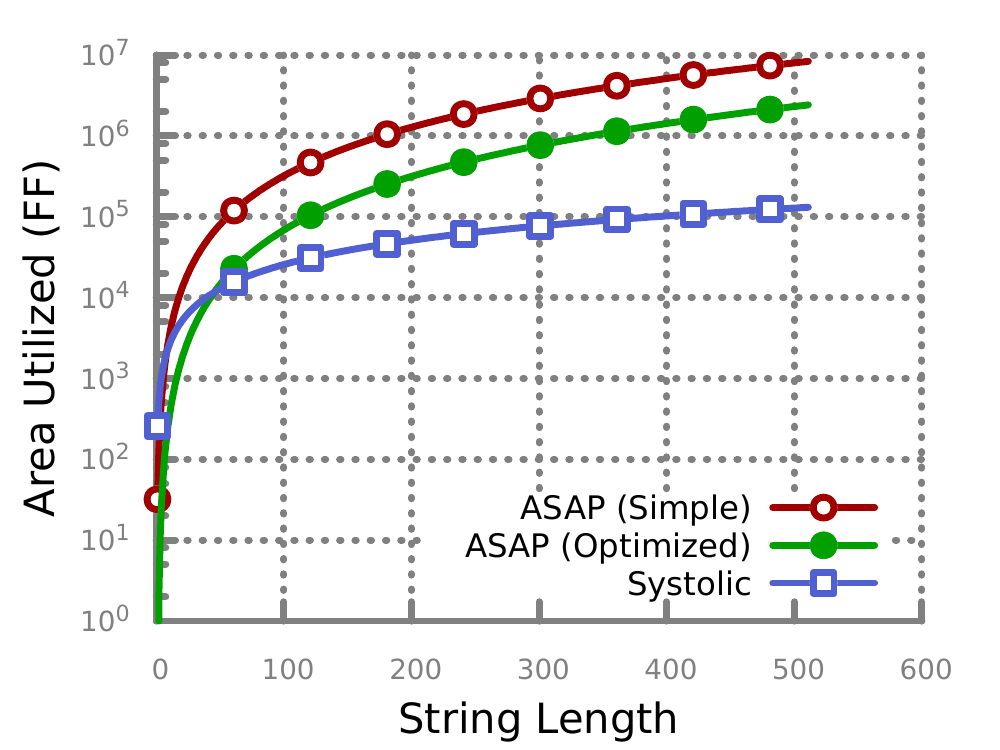}
    \caption{Scaling of FPGA resource utilization (accelerator size) with increase in input string 
        size.}
    \label{fig:asap_area}
\end{figure}

\subsection{Area-based Scaling} \label{sec:area_based_scaling}
The resource utilization of the ASAP accelerator scales quadratically with the lengths of the
sequences being compared. For example, \cref{fig:asap_area} shows the number of flip-flops used by
the ASAP accelerator with increasing string length, based on a 16$\times$16 square tile
size\footnote{This example does not include flip-flops required for the CAPI interface.}. In
comparison, an FPGA-based systolic array implementation of the LD computation~\cite{Lipton1985}
(described in \cref{sec:related_work}) scales linearly (i.e., $2N+1$, where $N$ is the length of the
strings being compared).  It is apparent that for larger sequences, ASAP quickly exhausts the amount
of available storage (flip-flops), even on the largest FPGAs available today. However, ASAP is able
to compute LD for shorter-read sequences (e.g., the 100-150 bp sequences that are typically obtained
from an Illumina HiSeq 2500) which are popularly being used in resequencing experiments. In
addition, we leave approximately $20\%$ of the area of the FPGA free, to allow the design compiler
to place-and-route the circuit without timing violations due to wiring delays.\footnote{There is no
simple analytical method to derive the optimal tile size, sequence size and free area on the FPGA,
as the synthesis tools are a black box.} As a result, we are able to fit a maximum 128 bp read
accelerator on our FPGA. Fitting larger blocks leads to timing violations because of delays
introduced by the on-chip interconnect. Given the industry trend towards FPGAs with larger
programmable area, in the future it should be possible to extend ASAP to read sequences that are
potentially thousands of nucleotides long. The results in the following sections show measurements
from a Power8 system of up to 128 bp inputs, and simulated results (using Xilinx Vivado's RTL
Simulator) for larger input sizes.

Currently, the ASAP accelerator can be used to compute LD for larger strings by adding a special
control algorithm in software to compute LD between sub-strings of the original queries, and combine
them to compute the result. The algorithm works by measuring (and storing) the time at which the
signal wavefront leaves the extremal DEs of the ASAP lattice, and reintroducing this signal
wavefront in the same lattice after updating the nucleotides to be another disjoint substring of the
queries. We leave the hardware implementation of this approach for future work.

\begin{table}[!t]
    \centering
    \label{speedup}
    \caption{Comparison of performance of end-to-end run-time for LD computation on CPU and ASAP 
        ($50^{th}$ percentile). Rows marked with ``*'' are simulated results for the FPGA.}
    \begin{tabular}{ccccc}
        \toprule
        \textbf{Read Size} & \textbf{CPU Baseline} & \textbf{ASAP} & \textbf{Speedup} \\
        \midrule
        64   & 1890 $\mu$s & 10.3 $\mu$s & 183$\times$ \\
        128  & 2083 $\mu$s & 10.7 $\mu$s & 194$\times$ \\
        192* & 3326 $\mu$s & 16.4 $\mu$s & 203$\times$ \\
        256* & 3906 $\mu$s & 17.2 $\mu$s & 219$\times$ \\
        320* & 4484 $\mu$s & 18.9 $\mu$s & 237$\times$ \\
        \bottomrule
    \end{tabular}
    \label{tab:asap_perf_cpu}
\end{table}

\subsection{Performance of the Accelerator} \label{sec:asap_perf}

The ASAP accelerator (configured in SW mode) is approximately $200\times$ faster than the
baseline C implementation of the SW algorithm for computing LD that is optimized to use single
instruction multiple data (SIMD; e.g., Intel AVX instructions) and simultaneous multi-threading
(SMT; e.g., Intel Hyperthreads) based multi-threading~\cite{Daily2016}. The baseline implementation
exploits inter-task parallelism (i.e., data parallelism) by processing multiple reads across
threads. \cref{tab:asap_perf_cpu} describes the comparison of the performance of a single lattice
ASAP accelerator. Having multiple cores on the CPU or multiple ASAP lattices on the FPGA does not
change this comparison, as each core/lattice is expected to be computing a separate unrelated
instance of the LD computation. The performance of ASAP depends not only on the size of the inputs,
but on the inputs themselves (i.e., more mismatched inputs mean a higher computation time). Hence we
present all ASAP measurements as the median across all the randomly generated reads. We observe that
a single ASAP lattice shows $\sim200\times$ speedup relative to a single CPU core
(containing 8 SMT threads and SIMD units), with potential improvements in performance with growing
input size (see \cref{tab:asap_perf_cpu}). Overall, a Power8 CPU chip contains six such
cores, whereas our implementation of ASAP can scale to four lattices (see \cref{fig:layout}).
Hence a chip-to-chip comparison yields a $133\times$ improvement in performance.

\begin{figure}[!t]
    \centering
    \subfigure[Input string length (Tile length $= 
    16$).]{\includegraphics[width=0.8\columnwidth]{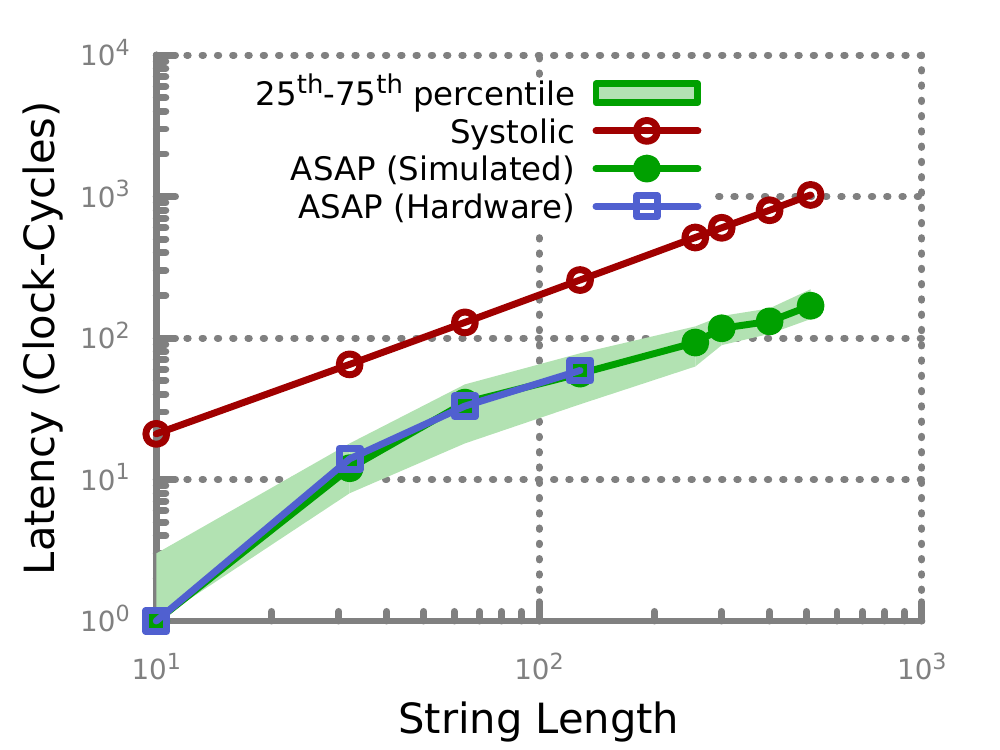}\label{fig:asap_perf_latency}}
    \subfigure[Tile length (Input length $= 
    128$).]{\includegraphics[width=0.8\columnwidth]{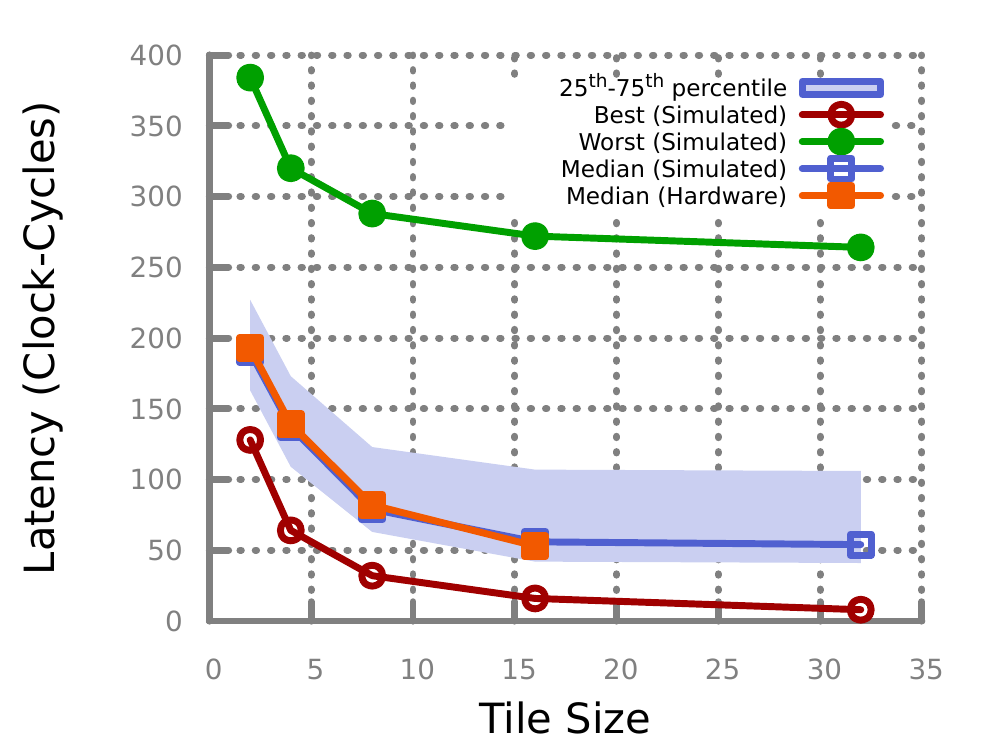} \label{fig:asap_perf_tile}}
    \caption{Latency of the accelerator as a function of the input string length. The shaded 
        area in both the graphs show $25^{th}$ and $75^{th}$ percentile measurement from 
        simulation.}
    \label{fig:asap_latency}
\end{figure}

\cref{fig:asap_perf_latency} illustrates the latency of the accelerator (without the overhead of
communication between the host and device) in computing LD (in the SW sense) for a single
read-reference pair. In contrast to traditional systolic-array-based accelerators, ASAP needs to
update only the cells (DEs) that can contribute to the LD computation (i.e., corresponding to the
colored cells in \cref{fig:wave_propagation_example}). Hence, throughput of the ASAP accelerator
can be computed in two ways: we can compute it either by considering the total number of cells in
the LD lattice, or by considering only the cells updated by ASAP. The first method which we refer to
as \emph{effective-GCUP/s} is directly comparable to traditional techniques as they too consider
updating all elements in the LD lattice. In terms of the first method, ASAP achieves an average of
$609.6$ GCUP/s ($10^9$ cell updates per second) for 128-bp reads; the second method, it achieves an
average of $204.8$ GCUP/s. This implies that in the median case, ASAP is approximately $5\times$
better than an equivalent systolic-array-based FPGA implementations (e.g., $122$ GCUP/s were
physically achieved on an FPGA in~\cite{Sirasao2015}\footnote{The comparison to~\cite{Sirasao2015}
is made based on numbers presented in their paper, and has not been re-implemented by us. Note that
the comparison is fair as the FPGAs in question are from the same architecture series as well as
running on similar clock frequencies.}). \cref{fig:asap_perf_tile} shows the effect of changing
tile-length on the latency of the accelerator. It is evident that there are diminishing returns for
increasing the tile length, with almost no improvement beyond tile size $16$.

Another point to note about \cref{fig:asap_latency} is that ASAP represents a method to
trade-off worst-case performance and average-case performance. The approximations that we present
may be slower than the baseline performance for the worst-case (i.e., when read mismatches reference
completely). However, we see that for representative data sets, the median performance as well as
the $75^{th}$ percentile performance are significantly better than the baseline. For the
short read alignment problem, we observe that matches occur more frequently than insertions,
deletions or mismatches. The ASAP accelerator can also be applied to other cases where insertions or
deletions are more frequent by dealing with those cases in combinational logic.

\subsection{Performance of the CAPI Interface} \label{sec:capi_perf}
\begin{figure}[!t]
    \centering
    \vspace{-.25cm}
    \subfigure[Mean observed 
    bandwidth.]{\includegraphics[width=0.8\columnwidth]{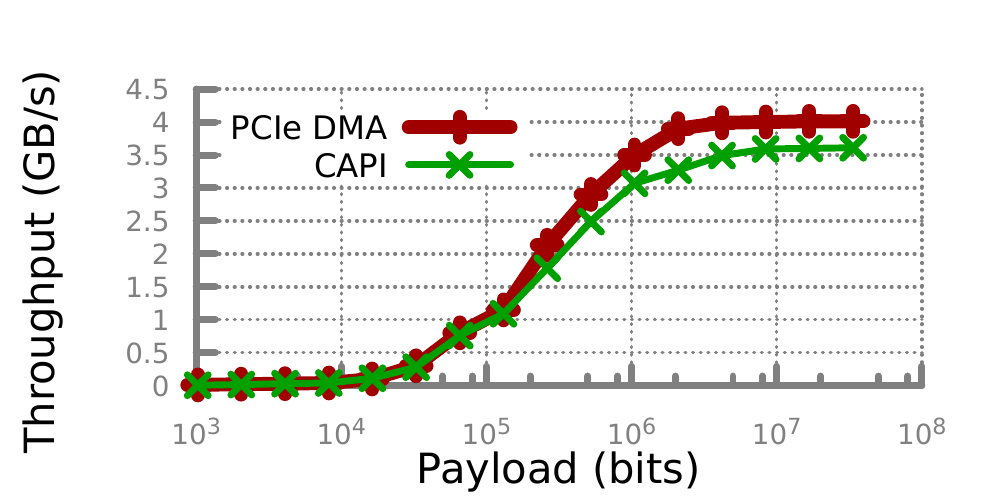}\label{fig:capi_bandwidth}}
    \subfigure[Fraction of cycles stalled due to unavailability of 
    data.]{\includegraphics[width=0.8\columnwidth]{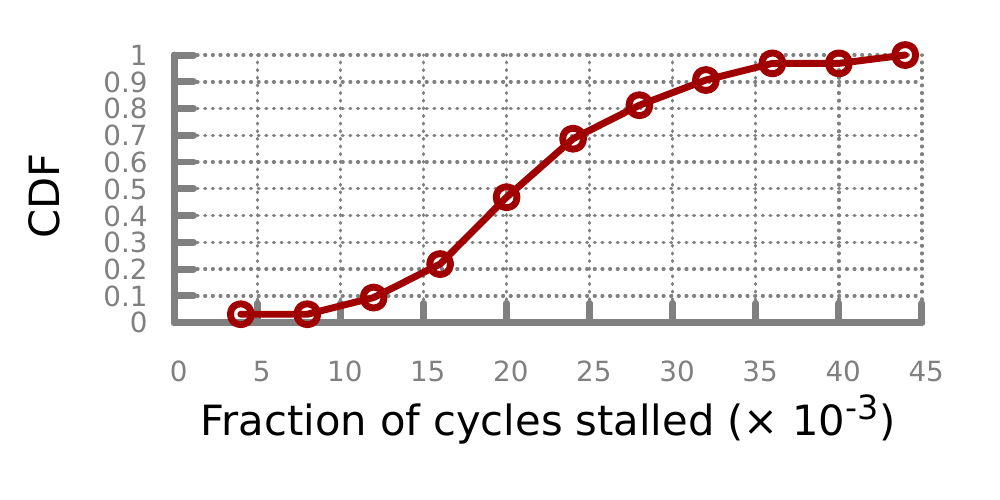}\label{fig:capi_stalls}}
    \caption{Mean host-accelerator bandwidth over the CAPI interface and its effect on the 
        performance of the ASAP accelerator.}
    \label{fig:capi}
\end{figure}

The ASAP accelerator benefits from the use of the CAPI interface, because CAPI  
\begin{enumerate*}
    \item significantly simplifies, and
    \item significantly streamlines
\end{enumerate*}
the process of initializing and communicating with the accelerator. We benefit from using a unified
virtual memory space across the PCIe bus with hardware-supported address translation, compared with
the traditional model, which requires significant hand-holding by an OS. For example, a typical
device driver would execute approximately $20k$ instructions, PCIe bounce-buffering, and
page-pinning to perform communication between host and accelerator. We performed measurements on the
CAPI interface using a loopback accelerator~\cite{Jaspers2015} (i.e., an accelerator reads a
cache-line and writes it back to a different location). We observed that (see
\cref{tab:capi_basic_performance} and \cref{fig:capi_bandwidth}) the CAPI interface can
perform random reads and writes with
\begin{enumerate*}
    \item sub-$\mu s$ latency, and
    \item $4$ GB/s bandwidth
\end{enumerate*}
which are both close to the measured native PCIe latency/bandwidth for the FPGA board used in the
evaluation. The one disadvantage that we observe with the CAPI interface is that it allows an AFU to
use only $50\%$ of the available peak-theoretical PCIe bandwidth. Our measurements of PCIe goodput
(i.e., bandwidth for user data to and from the accelerator) are similar to those from CAPI (see
\cref{fig:capi_bandwidth}). \footnote{We speculate that this limitation occurs because of
non-optimal interactions between the OS-modules (e.g., CAPI cache misses trigger TLB (ERAT in IBM
parlance) or page misses) and the PCIe-endpoint ASIC (e.g., dealing with out-of-order packet
delivery) on the FPGA board. We leave the optimization of such direct memory access (DMA) issues to
future work.} Bandwidth is currently not a limitation for the accelerator.
\cref{fig:capi_stalls} shows the fraction of the runtime of the accelerator spent in stall
over the execution of a large number of reads. However, moving to a larger FPGA that supports larger
ASAP lattices or multiple smaller ASAP lattices (executing in parallel), or clocking the ASAP
accelerator higher than 250 MHz will require larger bandwidth for the host-accelerator interface.

\begin{table}[!t]
    \centering
    \caption{Basic CAPI-based memory access performance for an AFU running on the Alpha-Data Board. 
        Latency measurements includes round-trip latency to shared memory as seen from the 
        accelerator.}
    \label{tab:capi_basic_performance}
    \begin{tabular}{ccp{3.5cm}c}
        \toprule
        \textbf{Interface} & \textbf{Payload (B)} & \textbf{Type} & \textbf{Measurement}\\
        \midrule
        PCIe & 128 & Mean read/write latency & $0.87~\mu s$ \\
        CAPI & 128 & Mean read/write latency & $126~ns$ \\
        CAPI & 128 & Mean read/write bandwidth & $3.88~GB/s$ \\
        \bottomrule
    \end{tabular}
\end{table}

\subsection{FPGA Resource Utilization} \label{sec:fpga_resource}
\begin{figure}[!t]
    \centering
    \subfigure[Scaling of FPGA resource utilization (accelerator size) with increase in number of 
    ASAP 
    lattices.]{\includegraphics[width=0.8\columnwidth]{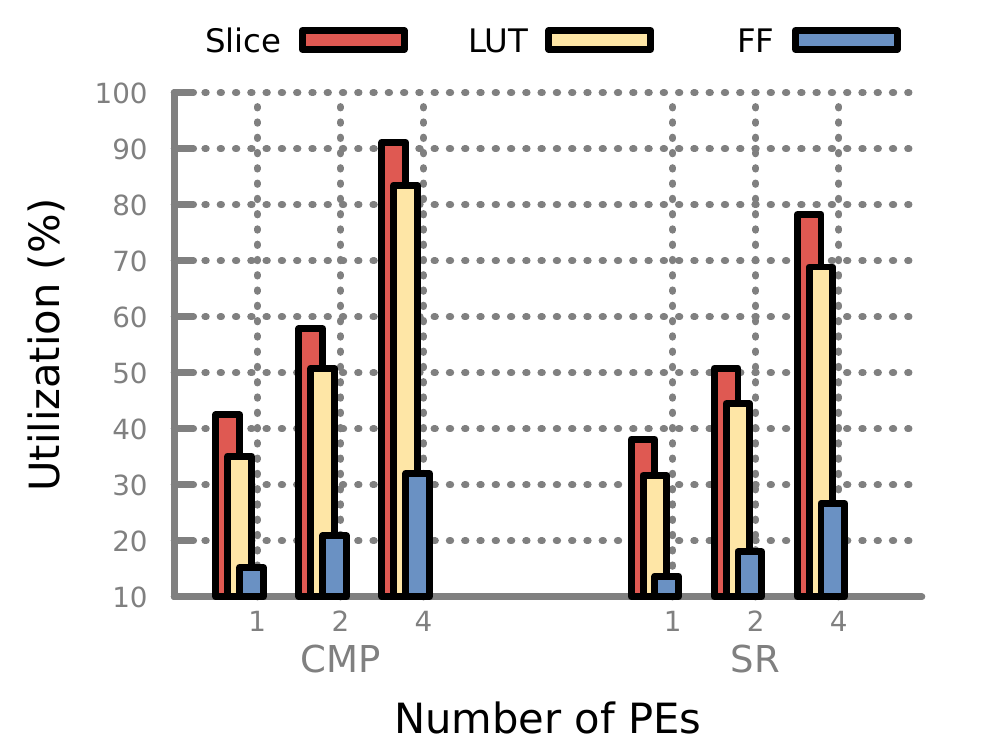}\label{fig:utilization}}
    \hspace{1em}
    \subfigure[Power dissipation from the ASAP accelerator with increase in number of ASAP lattices 
    per chip.]{\includegraphics[width=0.8\columnwidth]{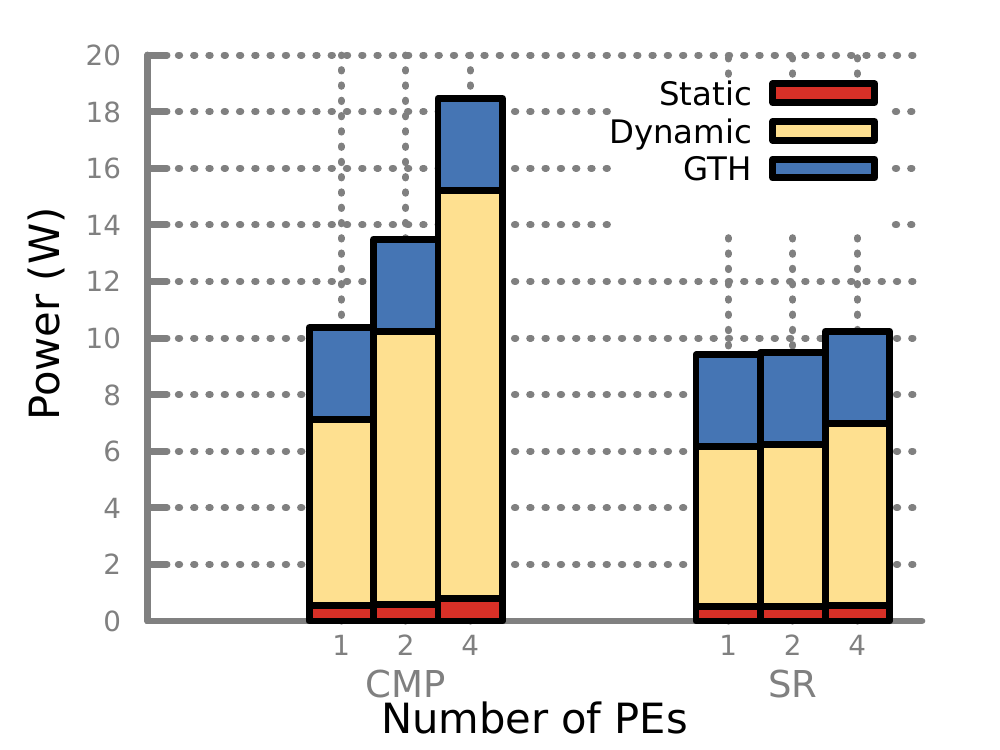}\label{fig:power}}
    \caption{Comparison of on-chip resource utilized by the \texttt{CMP} and \texttt{SR} 
        implementations of the ASAP design. Each ASAP lattice is $128\times 128$.}
    \label{fig:resources}
\end{figure}

Previously, \cref{sec:area_based_scaling} described the scaling behavior of the ASAP lattice 
with the input size; this section describes the overall on-chip resource utilization to implement 
the CAPI interface and multiple ASAP lattices on the FPGA. \cref{fig:resources} illustrates 
this utilization with the increasing number of lattices for two implementation styles for the ASAP 
delay element. First, the comparator based design that was presented in the original RaceLogic 
paper~\cite{Madhavan2014} (referred to as \texttt{CMP} in the figure), and second, the 
shift-register based design (presented in \cref{sec:asap}) that has been optimized for FPGAs 
(referred to as \texttt{SR} in the figure). \cref{fig:utilization} demonstrates the 
significant reduction (nearly $15\%$) in number of logic elements (i.e., slice resources) required 
to implement \texttt{SR} compared to \texttt{CMP}. This further translates to a $\sim 1.9\times$ 
reduction in power consumed by the \texttt{SR} design (shown in \cref{fig:power}). The proposed
design is nearly $18.8\times$ more power efficient than the IBM Power8 CPU ($\sim 10.1$ W compared
to $190$ W). This implies an overall $3,760\times$ ($= 200 \times 18.8$; based on
\cref{sec:asap_perf}) improvement over the CPU in performance/Watt terms.

Note that the power consumption for the chip is calculated from the synthesis tool (i.e., Xilinx
Vivado) and represents worst-case power consumed by the accelerator. However, the real power
consumption is input-dependent and lower than that mentioned above, as clock-gating on off-diagonal
delay elements will be enabled differently based on inputs (recall
\cref{fig:wave_propagation_example}). We computed this difference in power consumption using the
S824L's on-board power meters on the Flexible Service Processor (FSP).\footnote{The FSP is an
auxiliary processor on the S824L that is an always-on management processor enabling out-of-band
management of the server.} The FSP measurements report power consumption of the entire computer
system averaged over 30 s intervals. To calculate the power consumption of the ASAP accelerator, we
measured the difference in power consumed by the system when executing the 4-lattice instance of the
ASAP accelerator shown in \cref{fig:layout}, and when idling. We observed an average difference
(i.e., the ASAP accelerator's average power consumption) over 100 executions (of the entire
benchmark dataset) of 6.9 W with a standard deviation of 2.8 W. These measurements support our claim
that the actual power consumption of ASAP is lower than that reported by the synthesis tool.

\subsection{Integration into the SNAP Aligner} \label{sec:snap_perf}
We now compare the ASAP accelerator (configured in LV mode) when used in an end-to-end aligner (that
uses \cref{algo:basic_alignment_skeleton}), SNAP~\cite{Zaharia11}\footnote{We used version 1.0 of
the SNAP tool.}. We ensure that the maximum permissible LD in both the ASAP and SNAP implementations
of the LV algorithm are identical across all individual alignments. The baseline SNAP aligner
exploits parallelism in the alignment problem by dividing the work of aligning a set of reads among
all of the 192 threads available on the system. Since our current implementation of ASAP allows for
only one calling context on the host-side, we use ASAP in SNAP by maintaining a pool of memory
shared among all threads to communicate with the accelerator. The procedure for each thread
communicating with the accelerator is as follows:
\begin{enumerate*}
\item picks a read from the set it was assigned;
    \item queries the reference index for candidate locations for the read;
    \item contends for a lock, then writes nucleotides for the
    read and the candidate locations into shared memory;
    \item at this point, the accelerator reads from the shared memory and writes out the results to 
    another shared segment of memory; and
    \item polls for results from the accelerator using a test and test-and-set based locking protocol~\cite{Andrews2000}, then consumes the output.
\end{enumerate*}
This algorithm is another example scenario where CAPI is very beneficial, and we can make use
of cache coherence between the CPUs and FPGA to easily implement mutual exclusion.

\begin{figure}[!t]
    \centering
    \includegraphics[width=0.8\columnwidth]{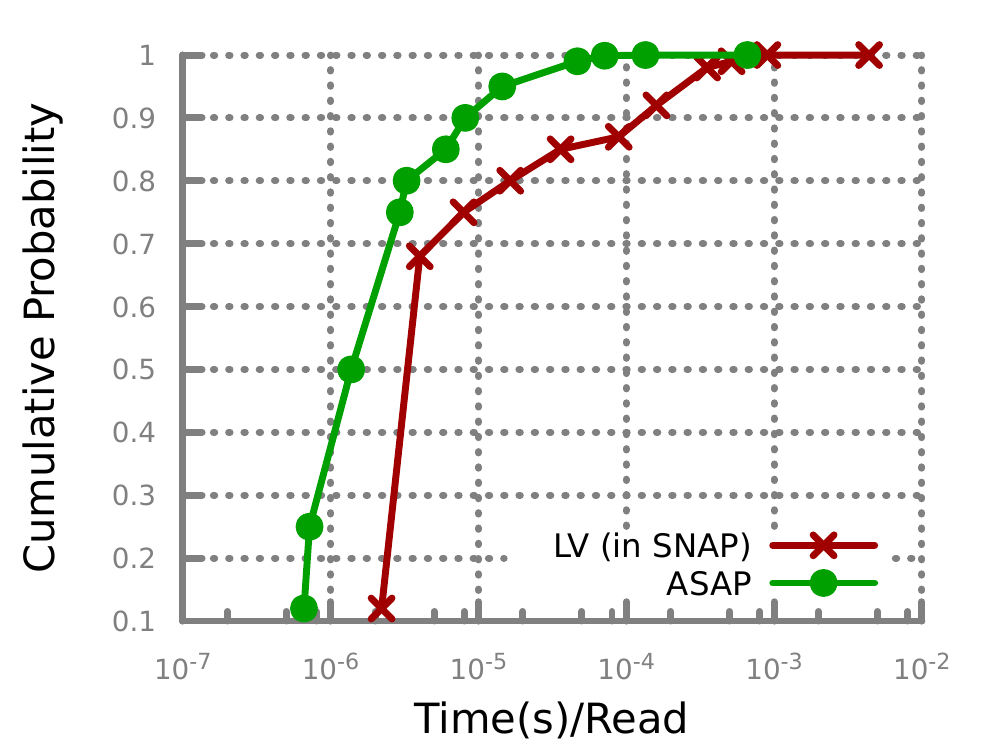}
    \caption{Comparing performance of SNAP using the default LV implementation and ASAP using the 
        SW algorithm.}
    \label{fig:asap_perf_snap}
\end{figure}

\cref{fig:asap_perf_snap} shows the time taken per read by the baseline and the ASAP
accelerator for all LD computations.  We see that there is a large spread for total time spent in
computing LD because some reads map to more regions of the reference than others. This variation is
an artifact of both the nature of the human genome and the read simulator's practice of picking
reads at random from the genome. The ASAP accelerator was used in SW mode in these experiments (a
quadratic complexity algorithm), but is compared to a CPU implementation of the LV algorithm (a
linear complexity algorithm). This comparison represents the most challenging acceleration use-case
for ASAP. We demonstrate how to configure ASAP to a LV accelerator in \cref{sec:asap_approximation}.
Overall, we see that the aligner is accelerated by $2\times$ (i.e., $\nicefrac{\text{1.85
hr}}{\text{0.92 hr}}$). This is close to the Amdahl's law limit of the SNAP algorithm based on our
measurements presented in \cref{tab:snap_callgraph}.

    \section{Related Work} \label{sec:related_work}
The sequence alignment problem has been addressed by an extensive body of work that looks at
algorithms and their high-performant implementations on CPUs and on accelerators like GPUs and
FPGAs. In this section we restrict ourselves to the comparison of the ASAP accelerator to other
implementations of the LD computations. Refer to \cref{sec:ld_and_alignment} for a discussion
of algorithms.

\textbf{On CPUs and GPUs.} The LD computation and sequence-alignment problem has been studied on
SIMD and MIMD processors that exploit parallelism in the problem at two levels. Inter-task
parallelism~\cite{Georganas2015} (using multiple cores to independently compute alignments of
different short reads), and intra-task parallelism~\cite{Farrar2006, Hughey1996} (using SIMD
instructions and efficient use of the memory hierarchy to effectively compute
\cref{eqn:basic_levenshtein_distance}). Most of the popular SW or NW implementations exploit
the use of both of these techniques. These techniques have also been applied to GPUs~\cite{Liu2006,
Liu2012, Zhao2014}. One such example is NVIDIA's NVBIO~\cite{NVBIO} library and the accompanying set
of tools nvBWT, nvFM-server. These look at accelerating the construction and look-up of data
structures that index the reference genome. The major disadvantages of this approach is the large
power consumption of these processors, and their restrictive lock-step parallelism based programming
models.

\textbf{On FPGAs and ASICs.} Custom hardware acceleration of the problem on FPGAs and ASICs has also
been widely studied. Most of the popular hardware architectures are based on systolic
arrays~\cite{Lipton1985, Hoang1993, Guccione2002, Zhang2007, Ahmed2015}. These architectures like
the SIMD and MIMD approaches, are limited by the amount of parallelism they can exploit. It has been
shown in \cite{Chen15}, that exploiting deeper pipelines with much larger inter-task parallelism can
potentially enable more efficient use of FPGAs. We may be able to use this optimization to further
increase the throughput of the accelerator, particularly on larger FPGAs that can sustain larger
off-chip bandwidth. Kaplan et al.~\cite{Kaplan2017} present an ASIC design for a
Processing-in-Memory accelerator for the SW algorithm that leverages resistive content-addressable
memory to compute matches/mismatches of nucleotides. ASAP represents a significant improvement
over~\cite{Kaplan2017} in throughput/Watt terms, i.e., ASAP achieves 61 GCUP/s/W
($=\nicefrac{609.6}{10.1}$) compared to their 53 GCUP/s/W. Turakhia et al.~\cite{Turakhia2018}
present an accelerator to perform long-read assembly, one step of which includes a SW-based
alignment (through a seed-and-extend approach). Aligning long reads (i.e. $\geq 1000$ base-pair
reads) poses a significantly different algorithmic challenge than aligning short-reads (i.e.,
$100-250$ base-pair reads) as the sequencing chemistry that produces the long reads are inherently
error prone. As a result,~\cite{Turakhia2018}'s SW implementation is not directly comparable to
ASAP. Alser et al.~\cite{Alser2017} present an FPGA based accelerator to efficiently filter
candidate locations to calculate LD. This accelerator is targeted at
Line~\ref{algo:line:candidate_location} of \cref{algo:basic_alignment_skeleton}, as opposed to ASAP
which targets Line~\ref{algo:line:edit_distance}, hence the accelerator can be used in addition to
ASAP to accelerate the end-to-end alignment process. More recent work~\cite{Rucci2018} has also
shown the benefit of distributing the compute intensive LD computation across multiple accelerators
(including CPUs, GPUs, FPGAs, Xeon Phis). We observe that ASAP significantly outperforms such
multi-accelerator systems both in terms of performance and performance per-Watt. The Host +
$2\times$ FPGA design presented in~\cite{Rucci2018} only achieves a 441.6 GCUP/s performance at 1.51
GCUP/s/W. In comparison ASAP achieves 609.6 effective GCUP/s at 61 GCUP/s/W on a single
FPGA.\footnote{The comparison is made across an equivalent generation of Altera and Xilinx FPGAs,
using \emph{effective-GCUP/s} (described in \cref{sec:asap_perf}).} Other work,
e.g.,~\cite{Peltenburg2016, Banerjee2017, Huang2017}, has demonstrated the use of
systolic-array-based designs to accelerate computations on Pair-HMM models, where gap-penalties are
replaced by probability distributions. That may be a future direction for the extension of the ASAP
design.

ASAP's design philosophy is most closely related to Madhavan et. al.'s
RaceLogic~\cite{Madhavan2014} ASIC design, which also encodes LD computations as circuit delay.
However, ASAP builds on this basic model to further optimize the design by using
\begin{enumerate*}
 \item approximation algorithms for the LD computation which maintains the total ordering of LDs, 
 and
 \item accelerating the most common computation (in this case the processing of ``matches'') in
 combinational circuitry thereby spending minimal runtime in its computation.
\end{enumerate*}
This is demonstrated by the fact that ASAP is $\sim 50 \times$ faster than a RaceLogic
implementation. Further, the nature of the alignment problem and the rapidly evolving sequencing
technology (i.e., read lengths), implies that fixed function ASICs are not favorable because of the
large monetary investment required and the inability of the accelerator to adapt to new input sizes.
ASAP circumvents these problems by using reconfigurable FPGAs. Of course, an ASIC  will almost 
always outperform an FPGA in energy efficiency because of its customized layout. Hence going 
forward, a design with a fixed function (i.e., ASIC-based) IO interface (i.e., CAPI) with a 
configurable substrate for ASAP accelerators might present an ideal trade-off.

\textbf{Comparison to Systolic Arrays.} Relative to the related work described above, ASAP has 
some decided advantages:
\begin{enumerate}[noitemsep,nolistsep,leftmargin=*]
    \item The systolic array based approaches require each element of the array to compute on as
    many bits as the maximum LD computed. Our approach requires only as many bits per delay element
    as the maximum delay between inputs at that point in the lattice.
    \item The earlier accelerators have to explore the entirety of the lattice before computing the
    LD. We show that the ASAP accelerator explores only the portions of the lattice that is
    reachable before the final result is produced. This represents a significant savings in run time
    and energy expended for computation.
    \item The ASAP accelerator can explore multiple elements in the lattice in under one clock 
    cycle by setting $\delta(\text{Match}) = 0$. Systolic array based architectures cannot perform 
    this optimization, as this creates large combinational chains which make timing closure 
    difficult to obtain.
\end{enumerate}

\textbf{On Neuromorphic Computers.}  Neuromorphic computing is modeled 
on biological neurons that communicate and compute using information encoded as voltage pulses, or 
spikes. Such spiking based models convey and process information via precise spike timing 
relationships measured across multiple communication paths. Though the computational model is 
similar in principle to the delay based computation outlined in this paper, realization of this 
technology, and its applications in domains other than pattern recognition is still an open 
research question~\cite{Merolla2014, Benjamin2014}. This field of research represents an avenue to 
extend ASAP using analog circuit components.
        
    \section{Conclusion and Future Work} \label{sec:conclusion}
This paper proposed ASAP, an accelerator for rapid computation of Levenshtein distance, in the
context of the short-read alignment problem. ASAP builds upon the idea that the LD between strings
can be approximated for the short-read alignment problem by encoding gap penalties in propagation
delays of circuit elements. We show that by effectively setting these delays, it is possible to
accelerate performance significantly, and at the same time ensure that the accuracy of alignment is
maintained. Accelerators like ASAP that synergize well with technologies like the CAPI interface
point to a new generation of HPC machines that will embrace heterogeneity and allow for efficient
handling of high throughput genomic data.

The ASAP accelerator, and the approach (based on heuristic approximations) presented in this paper,
can also be adapted to a variety of other problems in which a total ordering of LDs is computed. For
example, in signal processing, where different instances of a signal have to be aligned to compute
similarity~\cite{Levenshtein66}; in text retrieval, where misspelled words have to be accounted for
in a dictionary~\cite{Baeza-Yates1999}; and in virus- and intrusion-detection, where signatures have
to be aligned to a baseline~\cite{Kumar94}.

\textbf{Future Work.} Our future work will primarily look to extend ASAP to handle more complex
gap-penalty models. This paper describes the use of constant gap penalties (i.e., a fixed score is
assigned to every gap), which are commonly used in DNA alignment (e.g., in NCBI-BLASTN, or
WU-BLASTN~\cite{Sung2009}). We can extend ASAP to handle linear, affine, and convex gap penalties by
letting each DE track the propagation of the signal wavefront in the portion of the lattice before
it. Further, ASAP can be extended for use in the alignment of proteins by using substitution
matrices, like BLOSUM~\cite{Altschul1990}, which assign unique scores to each pair of residues.
    
    \section*{Acknowledgments}  \addcontentsline{toc}{section}{Acknowledgment}
This research was supported by several grants: in part by the National Science Foundation under
Grant No. CNS 13-37732; in part by the Blue Waters sustained-petascale computing project supported
by the National Science Foundation (awards OCI-0725070 and ACI-1238993) and the state of Illinois;
and in part by IBM Faculty Awards. We thank Jenny Applequist and Kathleen Atchley for
their help in preparing the manuscript.

    {
        \balance
        \bibliographystyle{IEEEtran}
        \bibliography{IEEEabrv,references}
    }
	\begin{IEEEbiographynophoto}{Subho S. Banerjee}
	is a PhD candidate in Computer Science at the University of Illinois at Urbana-Champaign. His
	research focuses on the design and implementation of workload optimized computing systems 
	(using hardware accelerator and parallel runtime environments) for data analytics workloads. He 
	holds a B.Tech. degree in Computer Science and Engineering from LNMIIT, India.
	\vspace{-1cm}
\end{IEEEbiographynophoto}
    
\begin{IEEEbiographynophoto}{Mohamed el-Hadedy}
	a is Research Scientist at the University of Illinois at Urbana-Champaign. He earned his 
	B.Sc and M.Sc degrees from the Mansoura University, Egypt in 2002 and 2006 respectively, and 
	his PhD degree in Electrical and Computer Engineering from the Telematics Department at the 
	Norwegian University of Science and	Technology, Trondheim, Norway in 2012. His main research 
	interests include FPGA-based accelerator design for Cryptography, Signal/Image Processing, 
	Robotics, and Genomics.
	\vspace{-1cm}
\end{IEEEbiographynophoto}
   
\begin{IEEEbiographynophoto}{Jong Bin Lim}
	received his B.S. degree in Electrical Engineering from University of Illinois at the 
	Urbana-Champaign in 2014. He is currently working toward his Ph.D degree in the department 
	of Electrical and Computer Engineering at the University of Illinois at Urbana-Champaign. 
	His current research interests include optimal System-On-Chip and accelerator design by using
	high-level synthesis, and hardware-software co-design.
	\vspace{-1cm}
\end{IEEEbiographynophoto}

\begin{IEEEbiographynophoto}{Zbigniew T. Kalbarczyk}
	is a Research Professor at the Electrical and Computer Engineering and the Coordinated Science Laboratory of the University of Illinois at Urbana-Champaign. Dr. Kalbarczyks research
	interests are in the area of design and validation of reliable and secure computing systems.
	\vspace{-1cm}
\end{IEEEbiographynophoto}

\begin{IEEEbiographynophoto}{Deming Chen}
	received the B.S. degree in computer science from the University of Pittsburgh, PA, USA, in 
	1995, and the M.S. and Ph.D. degrees in computer science from the University of California at 
	Los Angeles, in 2001 and 2005, respectively. He is a Professor with the ECE
	Department, University of Illinois at Urbana–Champaign, where he is the Donald Biggar Willett
	Faculty Scholar. His current research interests include system-level and high-level	synthesis,
	nano-systems design and nano-centric CAD techniques, GPU and reconfigurable computing, hardware
	security, and computational genomics.
	\vspace{-1cm}
\end{IEEEbiographynophoto}

\begin{IEEEbiographynophoto}{Steven S. Lumetta}
	received the A.B. degree in physics and the M.S. and Ph.D. degrees in computer
	science from the University of California, Berkeley, in 1991, 1994, and 1998, respectively.
	He is an Associate Professor of Electrical and Computer Engineering and a Research
	Associate Professor with the Coordinated Science Laboratory, University of Illinois at
	Urbana-Champaign. His research interests are in optical networking, high-performance networking 
	and computing, hierarchical systems, and parallel run-time software.
	\vspace{-1cm}
\end{IEEEbiographynophoto}

\begin{IEEEbiographynophoto}{Ravishankar K. Iyer}
	is the George and Ann Fisher Distinguished Professor of Engineering at the University of 
	Illinois at Urbana-Champaign. He holds appointments in the Department of Electrical and Computer
	Engineering, the Coordinated Science Laboratory (CSL), and the Department of Computer Science,
	serves as Chief Scientist of the Information Trust Institute, and is affiliate faculty of the
	National Center for Supercomputing Applications (NCSA).
\end{IEEEbiographynophoto}
\end{document}